\documentclass[notitlepage,twocolumn,prl,nobalancelastpage,amssymb,superscriptaddress,showpacs,longbibliography]{revtex4-2}

\usepackage{graphicx}
\usepackage{amsmath,amssymb,bbold,bm,color}
\usepackage{float}
\usepackage{epstopdf}
\usepackage{hyperref}
\usepackage{color, xcolor}
\usepackage{enumerate}
\usepackage{wasysym}
\usepackage{url}
\usepackage{dsfont}
\usepackage{braket}
\usepackage{comment} 
\usepackage{ulem}

\newcommand{\bd}{\mathbf{d}}
\newcommand{\be}{\mathbf{e}}
\newcommand{\bs}{\mathbf{s}}
\newcommand{\bB}{\mathbf{B}}
\newcommand{\bS}{\mathbf{S}}

\newcommand{\balpha}{\boldsymbol{\alpha}}

\newcommand{\bsigma}{\boldsymbol{\sigma}}

\newcommand{\cD}{{\cal D}}
\newcommand{\cC}{{\cal C}}

\definecolor{IEcolor}{RGB}{255, 153, 51}

\definecolor{FNcolor}{RGB}{200, 40, 200}

\renewcommand{\vec}[1]{{\bf{#1}}}

\hypersetup{colorlinks=true,linkcolor=blue,citecolor=blue,urlcolor=blue}

\begin{document}

\title{Quantum geometry and bounds on dissipation
in slowly driven quantum systems}

\author{Iliya Esin}
\affiliation{Department of Physics and Institute for Quantum Information and Matter, California Institute of Technology, Pasadena, California 91125, USA}
\address{Department of Physics, Bar-Ilan University, 52900, Ramat Gan, Israel}

\author{\'Etienne Lantagne-Hurtubise}
\affiliation{Department of Physics and Institute for Quantum Information and Matter, California Institute of Technology, Pasadena, California 91125, USA}
\affiliation{Département de Physique and Institut Quantique, Université de Sherbrooke, Sherbrooke, Québec, Canada J1K 2R1}

\author{Frederik Nathan}
\affiliation{Department of Physics and Institute for Quantum Information and Matter, California Institute of Technology, Pasadena, California 91125, USA}
\affiliation{Center for Quantum Devices and NNF Quantum Computing Programme, Niels Bohr Institute, University of Copenhagen, 2100 Copenhagen, Denmark}

\author{Gil Refael}
\affiliation{Department of Physics and Institute for Quantum Information and Matter, California Institute of Technology, Pasadena, California 91125, USA}

\date{\today}

\begin{abstract}
We show that  energy dissipation in slowly-driven, Markovian quantum
systems at low temperature is linked to the geometry of the driving protocol through the quantum (or Fubini-Study) metric. 
Utilizing these findings, we establish lower bounds on dissipation rates in two-tone protocols, such as those employed for topological frequency conversion. Notably, in appropriate limits these bounds are only determined by the topology of the protocol and an effective quality factor of the system-bath coupling. Our results bridge topological and geometric phenomena with energy dissipation
in open quantum systems, and further provide design principles for optimal driving protocols.
\end{abstract}

\maketitle

The geometry of quantum states~\cite{Xiao2010, Resta2011} is emerging as an important concept in condensed matter physics~\cite{Neupert2013, Kolodrubetz2013, Gao2014, Sodemann2015, Gao2015, Peotta2015,Julku2016,Liang2017, Bleu2018, Gao2019,Holder2020, Mitscherling2020, Rossi2021,Ahn2021, Takayoshi2021, Bhalla2022, Mitscherling2022,Trm2022, Lysne2023, Wang2023, Gao2023, Kaplan2024,Komissarov2024,Jankowski2025}. In particular, the {\it quantum} (or {\it Fubini-Study) metric}~\cite{Provost1980}---which defines a notion of distance on the manifold of quantum states---has been linked to various observable quantities including conductivity~\cite{Mitscherling2020, Mitscherling2022}, non-linear Hall effects~\cite{Gao2014, Sodemann2015, Wang2023, Gao2023}, optical responses~\cite{ Holder2020, Ahn2021, Takayoshi2021, Bhalla2022, Lysne2023} and superfluid stiffness~\cite{Peotta2015, Julku2016, Liang2017, Trm2022}. The renewed interest in the quantum metric provides a counterpoint to the much-studied effects of the Berry curvature in solids~\cite{Xiao2010}, as these two objects are obtained from the real and imaginary parts of a more general quantum geometric tensor.

Thermodynamics provides another context which hosts an emergent geometric description: namely, in quantum and classical systems a {\it dissipation metric} controls the energy dissipated by adiabatically traversing a path in parameter space~\cite{Avron2010, Avron2011, Avron2012, Sivak2012,Zulkowski2015,Ludovico2016,Bukov2019,Scandi2019,Abiuso2020,Bhandari2020,Loutchko2022,Alonso2022,Arrachea2023,Morimoto2023}. This geometric picture then provides a simple operating principle to minimize dissipation or entropy production for a given task, by following geodesics on the corresponding manifold.

\begin{figure}
  \centering
  \includegraphics[width=7.6cm]{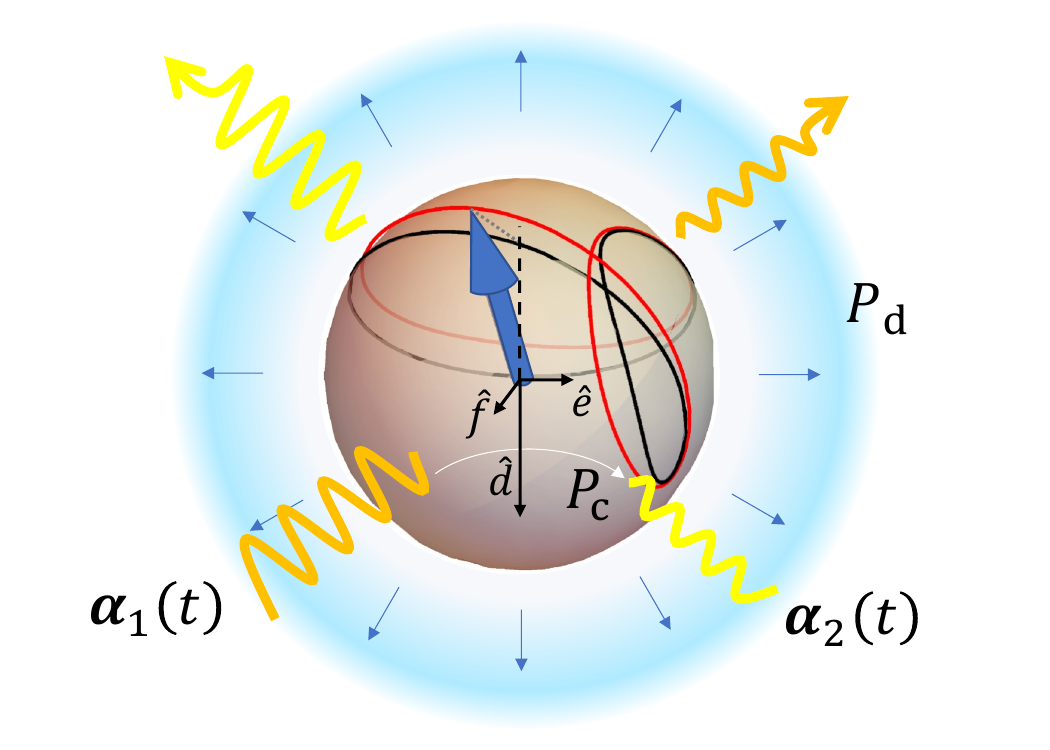}\\
  \caption{\textbf{The setup.} 
  A two-level system driven by the slowly-oscillating fields $\balpha_1$ and $\balpha_2$ and weakly coupled to a heat bath (represented by the blue halo). The black and red curves denote the trajectories of the Bloch-sphere vector $\bS_{0} = -\hat \bd$ representing the instantaneous ground state, and the steady-state vector $\bS_{\rm st}$ of the system, respectively. The instantaneous coordinate system consists of $\hat{\bd}$, its normalized time derivative $\hat{\vec e} = \dot{\hat \bd} / |\dot{\hat \bd}|$ and $\hat{\vec f} = \hat{\bd} \times \hat{\be}$. Leading diabatic corrections to the dynamics push the steady state in the $\hat{\vec f}$ direction, giving rise to energy pumping between $\balpha_1$ and $\balpha_2$ with rate $P_c$.  The lag between $\bS_{\rm st}$ and $\bS_{0}$ induced by the heat bath (along the $\hat{\be}$ direction) leads to geometric energy dissipation with rate $P_d$.}
   \label{fig:Setup}
\end{figure}

In this work we uncover a connection between these two distinct geometric notions. Considering slowly-driven quantum systems that are weakly coupled to a heat bath, we show that the low-temperature energy dissipation rate is controlled by geometric properties of the driving protocol through the  quantum metric. We then exploit inequality relations between the quantum metric and Berry curvature~\cite{Provost1980,Ozawa2021,Graf2021}
to establish various lower bounds on dissipation in a class of two-tone protocols~\cite{Martin2017} that support topological frequency conversion. 

{\it Setup.---} We first focus on
a two-level system subject to quasi-adiabatic driving, described by 
\begin{equation}
H(t) = \bd(t)\cdot \bsigma,
\end{equation}
where $\bsigma=(\sigma_x,\sigma_y,\sigma_z)$ is a vector of Pauli matrices spanning the Hilbert space.
The system is weakly coupled to a 
thermal bath at temperature $T_{\rm B}$~\cite{Leggett1987}. We consider the Markovian limit~\cite{Alicki1979,Blumel1991,Kohler1997,Breuer2000,Breuer_Petruccione,Hone2009,Avron2011,Mozgunov2020,Nathan2020, DiMeglio_2023},
where the coupling strength 
is weaker than the inverse correlation time of the bath, 1/$\tau_c$, 
of order $T_{\rm B}$ for generic (e.g. Ohmic) baths. The dynamics {of the density matrix $\rho(t)$} are then described by
a master equation of the form
\begin{equation}
    \dot\rho(t) = -i [H(t),\rho(t)]+\cD(t)\{\rho(t)\},
    \label{eq:MasterEquation}
\end{equation}
taking $\hbar=1$ throughout. Here, the dissipator $\cD$ is a linear superoperator that depends on the details of the system, the thermal bath, and the system-bath coupling.

We consider cases where the time dependence of the Hamiltonian, $H(t)$, is slow compared to $\tau_{\rm c}$.
In this regime, $\mathcal D(t)$ approaches the
dissipator resulting from a static Hamiltonian---namely, one that relaxes the system towards instantaneous thermal equilibrium with the bath~\cite{Breuer_Petruccione,Nathan2020,DiMeglio_2023}, captured by $\rho_{\rm eq}(t) =  e^{-\beta H(t)}/{\rm Tr}[e^{-\beta H(t)}]$ with $\beta = 1/T_{\rm B}$. However, the time dependence of  Eq.~\eqref{eq:MasterEquation} implies that its steady-state solution 
generally differs from $\rho_{\rm eq}(t)$ due to diabatic corrections. As we describe in this work, such corrections cause geometric dissipation.

It is convenient to
formally reexpress Eq.~\eqref{eq:MasterEquation} as a Bloch equation~\cite{Bloch1946} for the vector
$\vec S(t)={\rm Tr}[\rho(t)\cdot \bsigma]$,
\begin{equation}
\dot {\vec S}(t) = 2 \bd(t)\times \vec S(t)-\Gamma(t) [\vec S(t)-\vec S_0(t)].
\label{eq:BlochEquation}
\end{equation}
Here the dissipator is reparametrized via the relaxation matrix $\Gamma_{ij}(t) = -\frac12{\rm Tr} \left[ \sigma_i \cD(t)\{\sigma_j\} \right]$, and $\bS_0= -\tanh(\beta\Delta/2) \hat{\bd}$ denotes the Bloch vector corresponding to $\rho_{\rm eq}(t)$~\footnote{For a general dissipator $\cD(t)$, $\bS_0(t) $ can be defined from $\cD$ as $\bS_0=\frac{\Gamma^{-1}(t)}{2}{\rm Tr}[\bsigma \cD(t)\{1\}]$.},  with $\Delta=2|\bd|$ the spectral gap and $\hat{\bd}\equiv \bd /|\bd|$.
Focusing on the low-temperature limit 
where quantum geometric effects dominate, we set $\bS_0 = - \hat{\bd}$. Finite temperature corrections are discussed in the SM.

We consider  
dissipators that are isotropic with respect to
$H(t)$. In this case, the most general form of the
relaxation matrix~\cite{DiMeglio_2023} reads
\begin{equation}
    \Gamma_{ij}=\frac{1}{\tau_1}\hat\bd_i\hat\bd_j + \frac{1}{\tau_2} \left( \delta_{ij}-\hat\bd_i\hat\bd_j \right) + \delta \epsilon_{ijk} \hat{\bd}_k,
    \label{eq:Gamma_isotropic}
\end{equation} 
where $\tau_1$, $\tau_2$ are scalars that can be interpreted as longitudinal and transversal relaxation times in the instantaneous eigenbasis of the Hamiltonian, with $\tau_2 \leq 2 \tau_1$~\cite{levitt2008}. In the Supplementary Material (SM)~\footnote[100]{See Supplemental Material for details.} we derive  Eqs.~\eqref{eq:MasterEquation}-\eqref{eq:Gamma_isotropic}, with an explicit form of $\Gamma$, from a {concrete} microscopic 
model (see also Ref.~\cite{DiMeglio_2023}). The last term in Eq.~\eqref{eq:Gamma_isotropic} 
can be absorbed in Eq.~\eqref{eq:BlochEquation} by rescaling  the vector $\bd(t)$---we therefore set $\delta=0$ for simplicity.

{\it Low-temperature steady state and dissipation.---}A steady-state solution to Eq.~\eqref{eq:BlochEquation} can be found, up to leading order in 
diabatic corrections, by performing a rotating frame transformation~\footnotemark[100],
\begin{equation}
    \bS_{\rm st} = \frac{{- \left( 1+ \Delta^2 \tau_2^2 \right)\hat{\bd} + \tau_2 \dot{\hat{\bd}} + \Delta \tau_2^2  \left( \hat{\bd} \times \dot{\hat{\bd}} \right) }}{1 + \Delta^2 \tau_2^2 + \tau_1 \tau_2 |\dot{\hat{\bd}}|^2}
    + {\cal O} \left(\frac{|\dot{\hat \bd}|^2}{\Delta^2} \right).
    \label{eq:inst_ST}
\end{equation}
This result can be visualized with the help of Fig.~\ref{fig:Setup}. As the Hamiltonian varies in time, the direction of the Bloch vector $\hat{\bd}$ changes. When driving is quasi-adiabatic, the steady state follows the trajectory of $- \hat{\bd}(t)$ to remain close to the instantaneous ground state. Dissipation to the heat bath causes the spin to lag behind the Hamiltonian by the contribution $\propto \dot{\hat{\bd}}$ in Eq.~\eqref{eq:inst_ST}. There is also a diabatic correction to the coherent dynamics, {due to the precession of the spin,} that pushes the steady state in the direction perpendicular to $\hat{\bd}$ and $\dot{\hat{\bd}}$, leading to the Berry curvature as discussed below.

In the steady state, the (long-time) average rate of energy dissipation, $\overline P_d$, equals the average rate of energy transfer into the system from the drive: $\overline P_d = \int_0^T \frac{dt}{T} \langle\dot H(t)\rangle $, where $\langle\dot H(t)\rangle=\dot \bd\cdot \bS_{\rm st}$. Inserting $\bS_{\rm st}$ from Eq.~\eqref{eq:inst_ST}, to leading diabatic order we find
\begin{equation}
\overline P_{\rm d}
= \int_0^T \frac{dt}{T}  \left[ \frac{1}{2} \frac{\Delta \tau_2 |\dot{\hat{\bd}}|^2 - \dot{\Delta} ( 1 + \Delta^2 \tau_2^2)}{1  + \Delta^2 \tau_2^2 +\tau_1 \tau_2|\dot{\hat{\bd}}|^2 }
\right] .
\label{eq:Wd_steady_state}
\end{equation}
The dependence of $\overline P_{\rm d}$ on $\tau_2$ may seem surprising, as dissipation in time-independent systems is set only by $\tau_1$. Its appearance is a consequence of the fact that $\tau_2$ controls dephasing in the {\it instantaneous eigenbasis} of the Hamiltonian, while the steady-state Bloch vector $\bS_{\rm st}$ lags behind
$\bS_{0} = -\hat {\bd}$, as shown in Eq.~\eqref{eq:inst_ST} and Fig.~\ref{fig:Setup}. Note that $\tau_1$ still constrains dissipation: 
$\overline P_{\rm d}$ appropriately vanishes when $\tau_1 \rightarrow \infty$.

Focusing for the rest of this work on the typical case where $\tau_1 \sim \tau_2$~\footnote{This condition can be extended to  $\tau_2/\tau_1 \gg |\dot{\hat{\bd}}|^2/\Delta^2$},
Eq.~\eqref{eq:Wd_steady_state} can be simplified by neglecting the higher-order diabatic contribution $\sim |\dot{\hat \bd}|^2$ in the denominator. 
{In this case the term proportional to $\dot \Delta$ vanishes when integrating over a period, as it can be rewritten as a total time derivative. We thus obtain}
\begin{equation}
    \overline P_{\rm d} 
= \int_0^T \frac{dt}{T}  \left[ \frac14\gamma |\dot{\hat\bd}|^2 \right],
\label{eq:Pd_simplified}
\end{equation}
where $\gamma=\frac{2 \Delta \tau_2}{1 + \Delta^2 \tau_2^2}$. In the limit of weak relaxation $\Delta \tau_2 \gg1$, $\gamma$ is proportional to the number of precession cycles within the relaxation time $\tau_2$, and can thus be understood as an effective {inverse} quality factor.

\begin{figure}
  \centering
  \includegraphics[width=8.6cm]{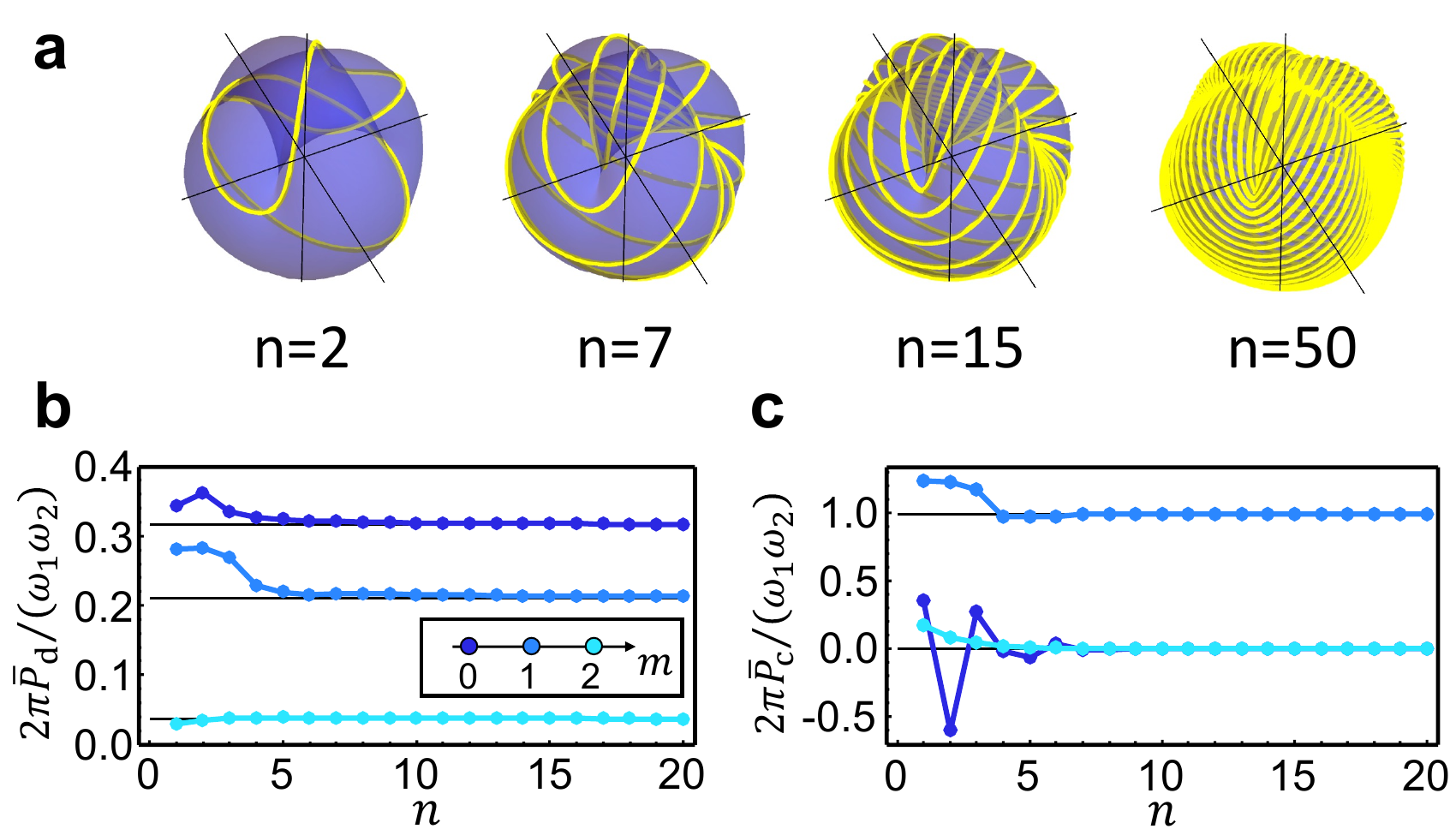}\\
  \caption{\textbf{Approaching the incommensurate limit.} \textbf{a.} Trajectories on the surface $\bs = \frac{1}{2} \sqrt{\gamma} \hat{\bd}$, for different values of $\omega_2/\omega_1=\frac{n}{n+1}$.
  \textbf{b.} Average dissipation rate along the ``Lissajous curves'' as a function of $n$. We use the driven spin model in Eq.~\eqref{eq:spin_Hamiltonian} and consider $b_{11}=b_{22}=1$, $b_{12}=b_{21}=0.5$, $\theta=0$, $\tau_2=10$, and three values of $m$ indicated in the inset. \textbf{c.} Average frequency conversion rates as a function of $n$ for the same parameters as in {\textbf b}, which quickly converge to the quantized result expected for incommensurate driving, Eq.~\eqref{eq:ChernNumber}.
   \label{fig:Commensurate}}
\end{figure}

{\it Geometric interpretation.---} Let us assume that
$\bd$ is controlled  by a set of $M$ parameters $(\alpha^1,\ldots, \alpha^M)=\balpha$ that are slowly changed over time:
$\bd(t)=\bd(\balpha(t))$. The average dissipation in Eq.~\eqref{eq:Pd_simplified} can then be written as an integral of a dissipation metric $\Lambda$ 
over the path traced by time evolution,
\begin{equation}
    \overline P_{\rm d} = \int_0^T \frac{dt}{T} \Lambda_{ij}\dot\alpha^i \dot\alpha^j, 
    \label{eq:Wd_omegas}
\end{equation}    
where
\begin{equation}
\Lambda_{ij}=\gamma G_{ij} ~ , ~ G_{ij}= \frac{1}{4} \left( \partial_{\alpha^i} \hat{\bd} \cdot \partial_{\alpha^j} \hat{\bd} \right).
\label{eq:a_tale_of_two_metrics}
\end{equation}
Here $G$ is the quantum metric in the parameter space describing the driving fields, $i$ and $j$ label coordinates in $M$-dimensional space and
Einstein summation convention is assumed throughout.
This constitutes the main result of our work, and connects to earlier works on quantum thermodynamics~\cite{Sivak2012,Zulkowski2015,Scandi2019, Abiuso2020,Bhandari2020,Loutchko2022,Alonso2022,Arrachea2023} that
provide an operating principle to minimize dissipation for a given task: the trajectory of
$\balpha(t)$ should follow a geodesic 
at uniform speed on the surface defined by $\Lambda$. 
When $\gamma$ is {nearly} constant
this surface is described by $\bs = \frac12\sqrt{\gamma} \hat{\bd}$, such that $\Lambda_{ij} = \partial_{\alpha^i} \bs \cdot \partial_{\alpha^j} \bs$, see Fig.~\ref{fig:Commensurate}a. 

We can also generalize Eq.~\eqref{eq:a_tale_of_two_metrics} for $N$-level systems~\footnote[100]{See Supplemental Material for details.}. Here the low-temperature dissipation metric $\Lambda^{(N)}$ instead involves a more complicated combination of quantum geometric and spectral quantities. Importantly,
$\Lambda^{(N)}$ can be bounded from above and below in terms of the ground-state quantum metric, $G_{ij}^{(N)} = \frac12{\rm Tr}[\partial_{\alpha^i} P_0 \partial_{\alpha^j} P_0]$, with $P_0(\balpha)$ the projector into the ground state of $H(\balpha)$:
\begin{equation}
G^{(N)} \succeq 
\Lambda^{(N)} \succeq \gamma^{(N)} G^{(N)}.
\end{equation}
Here $A\succeq B $ implies that $A-B$ is positive semidefinite, and 
$\gamma^{(N)} = \min_n \frac{2 \Delta_{n} \tau_{0n}}{1 + \Delta_{n}^2 \tau_{0n}^2}$, where $\Delta_{n}$ is the gap to the $n$-th excited state and $\tau_{0n}$ the relaxation 
time between the $n$-th exicted state and the ground state.

{\it Application to topological frequency conversion.---} The connection between dissipation and quantum geometry uncovered above hints at a possible role of topology. To elucidate this role, we now specialize to time evolution characteristic of topological frequency conversion protocols~\cite{Martin2017}, where $\balpha(t) = \balpha_1(t) + \balpha_2(t)$ describes two harmonic drives
with frequencies
$\omega_1$ and $\omega_2$, respectively (see Fig.~\ref{fig:Setup}). Here, the quantum metric in Eqs.~\eqref{eq:Wd_omegas}, \eqref{eq:a_tale_of_two_metrics} can be re-expressed as a function 
of the two phases $\phi_a = \omega_a t$ (with $a = 1,2$) describing the drives, {reading} $g_{ab} = G_{ij} \frac{\partial \alpha^i}{\partial{\phi_a}}\frac{\partial \alpha^j}{\partial{\phi_b}}$. The dissipation rate is then given by
\begin{equation}
\overline P_{\rm d} 
= \omega_a \omega_b  \int_0^T \frac{dt}{T} \gamma g_{ab}.
\label{eq:DissipationRate}
\end{equation}
Similarly, the transferred power {from mode $1$ to mode $2$} in the steady state can be calculated through $P_{12} = \frac{1}{2} \left( \frac{\partial \bd}{\partial \phi_1} \dot \phi_1 - \frac{\partial \bd}{\partial \phi_2} \dot \phi_2 \right) \cdot \bS_{\rm st}$.  This term is anti-symmetric in indices $1$ and $2$, and thus does not contribute to the net energy dissipation. Using the steady-state vector in Eq.~\eqref{eq:inst_ST}, one obtains $P_{12} = \gamma \left[(\dot \phi_1 )^2g_{11}  - ( \dot \phi_2)^2g_{22} \right] - P_{\rm c}$. The first two terms denote the difference in energy dissipation due to each drive,
and the frequency conversion power $P_c$ averages to
\begin{equation}
    \overline{P}_{\rm c} = \omega_1 \omega_2 \int_0^T \frac{dt}{T} \left[ \frac{\tau_2^2\Delta^2}{1+\tau_2^2\Delta^2}
    \right] \Omega_{12},
\label{eq:TFC_Rate}
\end{equation}
where $\Omega_{12}=\frac 12 \hat \bd \cdot \left( \partial_{\phi_1} \hat \bd \times \partial_{\phi_2} \hat \bd \right)$ is the Berry curvature associated with the phases of the two drives. 

\begin{figure}
  \centering
  \includegraphics[width=8.6cm]{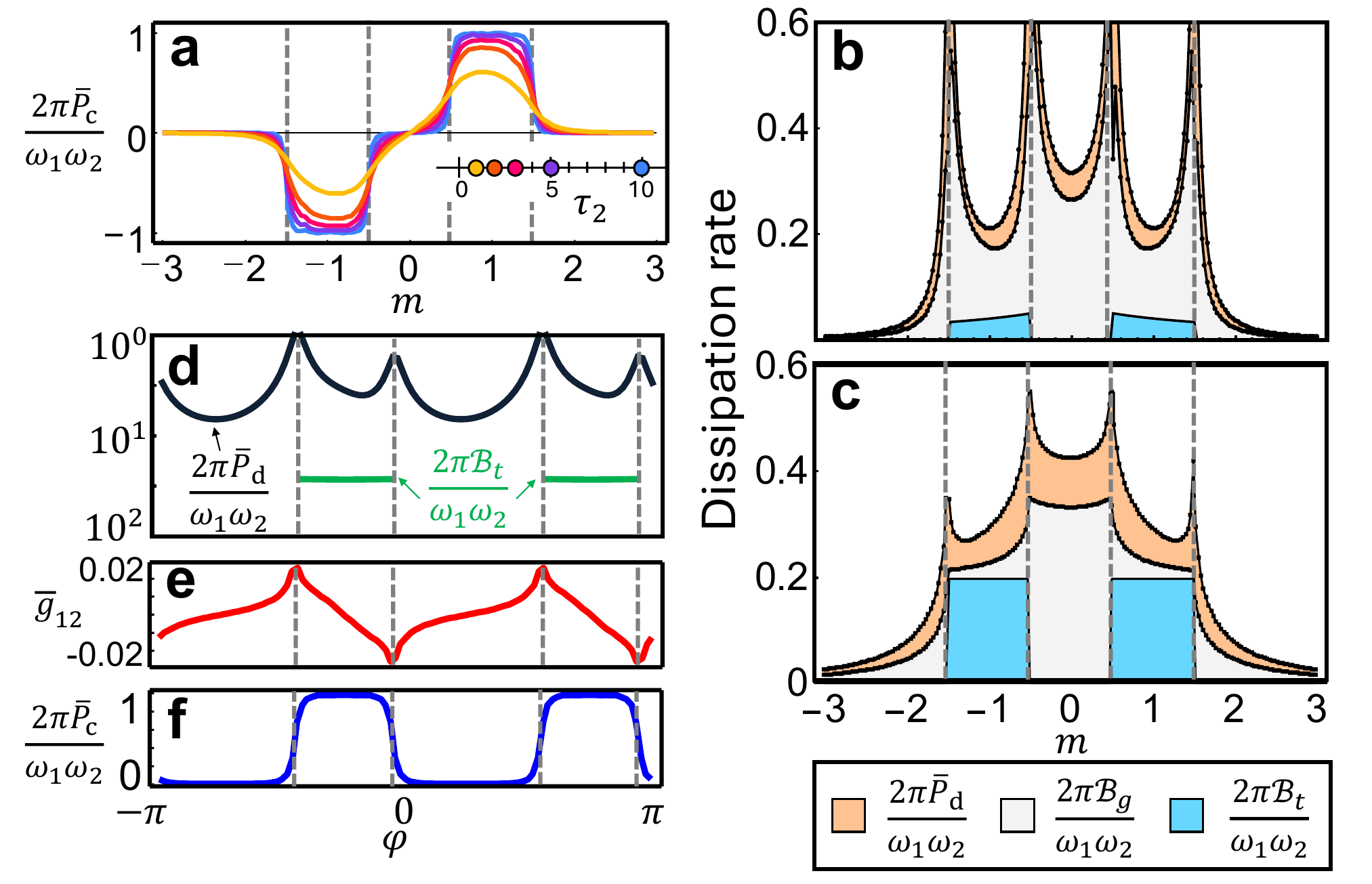}\\
  \caption{
\textbf{Frequency conversion and bounds on dissipation.} 
(\textbf{a}) Frequency conversion rate in the driven spin model, Eq.~\eqref{eq:spin_Hamiltonian},
as a function of $m$ for $\theta=0$ and  different values of $\tau_2$ indicated in the inset. In the limit $\tau_2\to \infty$, $\overline{P}_{\rm c}\to \frac{\omega_1 \omega_2}{2 \pi} \cC$, see Eq.~\eqref{eq:ChernNumber}. 
(\textbf{b})
Energy dissipation rate compared to the bounds 
in Eqs.~\eqref{eq:geometric_bound} and \eqref{eq:BoundSymmetric}, 
using the same parameters as in {\textbf a} and $\tau_2=10$. 
(\textbf{c}) {\it Idem} for the constant-gap Hamiltonian with
$\Delta_0=1$. 
(\textbf{d}) 
Dissipation rate and the topological bound for $\theta=0.2\pi$, $m=1.2$, and $\tau_2=10$ as a function of $\varphi$. 
(\textbf{e}) The value of $\overline g_{12}=\oiint g_{12} d\phi_1 d\phi_2$ and (\textbf{f}) the frequency conversion rate, for the same parameters as in {\textbf d}. We use $b_{11}=b_{22}=1$ and $b_{12}=b_{21}=0.5$ throughout.
   \label{fig:bounds}}
\end{figure}

When the drive frequencies $\omega_1$ and $\omega_2$ are incommensurate,
the system explores its entire phase space in the long-time limit. The 
 averages in Eqs.~\eqref{eq:DissipationRate} and \eqref{eq:TFC_Rate} can then be replaced by an average over the phases $\phi_a=\omega_a t$,
$\int_0^T \frac{dt}{T} \to \oiint \frac{d\phi_1 d\phi_2}{4 \pi^2}$. In the limit of weak relaxation $\Delta \tau_2 \gg 1$,
the frequency conversion rate acquires a clear topological character, $\overline P_{\rm c}=\frac{\omega_1\omega_2}{2\pi} \cC$, where 
\begin{equation}
    \cC=\frac{1}{2\pi}\oiint d\phi_1d\phi_2 \Omega_{12}
    \label{eq:ChernNumber}
\end{equation} 
is the corresponding Chern number, see Fig.~\ref{fig:bounds}a. In this regime, the two-level system transfers the energy of $\cC$ photons of mode $1$ per period of mode $2$ (or vice versa) between the two drives~\cite{Martin2017}.

For commensurate frequencies $\omega_1$ and $\omega_2$, the trajectory of $\balpha(t)$ 
on the Bloch sphere is a ``Lissajous'' curve~\cite{Nathan2022}. For a given ratio $\omega_2/\omega_1= \frac{n}{n+1}$, with $n\in \mathds{N}$, the time 
to complete one cycle increases as $T=(n+1)\frac{2\pi}{\omega_1}$, see Fig.~\ref{fig:Commensurate}a.
We compute the dissipation and frequency conversion rates using Eqs.~\eqref{eq:DissipationRate} and \eqref{eq:TFC_Rate}; the data are plotted in Figs.~\ref{fig:Commensurate}b,c. Both quantities rapidly converge for $n\gtrsim 5$ to their respective asymptotic values ($n\to \infty$) 
 corresponding to incommensurate driving.

{\it Bounds on dissipation and topology.---}We now use the connection between energy dissipation and the quantum metric
to derive lower bounds on dissipation for topological driving protocols. 
We consider a simple scenario in which the system and drives are in a symmetric configuration with $\overline{g}_{12} \equiv \oiint   d\phi_1d\phi_2 g_{12} = 0$
{(see SM~\footnote[100]{See Supplemental Material for details.}  for an analysis of the general case where  $\overline{g}_{12} \neq 0$.)}
In this case, the average dissipation in Eq.~\eqref{eq:DissipationRate} arises only from the diagonal components of the quantum metric, proportional to $\omega_1^2 g_{11}+\omega_2^2 g_{22}$. These two terms can be bounded from below by $2\omega_1\omega_2\sqrt{g_{11} g_{22}}  \geq 2\omega_1\omega_2\sqrt{ \det g}$.
Next, we use the identity $\sqrt{ \det g} \geq |\Omega_{12}|/2$ (saturated for two-band models~\cite{Ozawa2021}), leading to a geometric lower bound on dissipation, $\overline P_{\rm d} \geq \mathcal{B}_{\rm g}$, where
\begin{equation}
\mathcal{B}_{\rm g} = \omega_1 \omega_2 \oiint \frac{d\phi_1 d\phi_2}{4 \pi^2}\gamma | \Omega_{12}|.
\label{eq:geometric_bound}
\end{equation}
The geometric bound is saturated when $\omega_1 \to \omega_2$ and the trace condition inequality~\cite{Roy2014} is saturated,  ${\rm Tr} [g] = |\Omega_{12}|$.

A topological lower bound, $\mathcal{B}_{\rm t} \le \mathcal{B}_{\rm g}$, can be obtained by replacing the 
term $\gamma(\phi_1,\phi_2)$ by its minimal value $\gamma_{\rm min}=\min_{\phi_1,\phi_2}\gamma(\phi_1,\phi_2)$, yielding
\begin{equation}
\mathcal{B}_{\rm t}=  \frac{\omega_1 \omega_2}{2 \pi} \gamma_{\rm min}|\cC|.
 \label{eq:BoundSymmetric}
 \end{equation}
It follows from Eq.~\eqref{eq:BoundSymmetric} that a finite frequency conversion rate ($|\cC|>0$) implies a minimum dissipation rate in the symmetric drive configuration. Further, $\mathcal{B}_{\rm g}$ in Eq.~\eqref{eq:geometric_bound} provides a lower bound on dissipation for topologically-trivial protocols, with $\cC=0$, yet with
a non-vanishing Berry curvature distribution. 

{\it Driven spin system.---}To illustrate our results in a concrete physical setting, we consider a spin driven by two elliptically-polarized magnetic fields~\cite{Martin2017},
$\bB_1(t) =\left( 0,b_{11}\sin \omega_1 t,b_{12}\cos \omega_1 t \right)$ and $
    \bB_2(t) = \left( b_{21}\sin \omega_2 t,0,b_{22}\cos \omega_2 t \right)$.
The Hamiltonian reads
\begin{equation}
    H(t)=\vec m \cdot  \bsigma + \left( \bB_1(t) + \bB_2(t) \right) \cdot\bsigma,
    \label{eq:spin_Hamiltonian}
\end{equation}
where $\vec m=m(\sin \theta\cos\varphi,\sin \theta \sin\varphi,\cos\theta)$ is a time-independent Zeeman coupling. 
This model exhibits distinct topological regimes as a function of $\vec m$, characterized by $\cC = \{-1, 0, 1\}$ (see Fig.~\ref{fig:bounds}a).
For simplicity we
take $\tau_2$ to be time independent.

Figure~\ref{fig:bounds}b shows the dissipation in the different topological regimes of Eq.~\eqref{eq:spin_Hamiltonian}, calculated numerically using Eq.~\eqref{eq:DissipationRate} for $\theta=0$. 
 For this value of $\theta$, the symmetric drive condition is satisfied, $\overline g_{12} = \oiint  g_{12}d\phi_1d\phi_2=0$; the energy dissipation rate is therefore bounded by $\mathcal{B}_{\rm g}$ and $\mathcal{B}_{\rm t}$ [Eqs.~\eqref{eq:geometric_bound} and \eqref{eq:BoundSymmetric}].
 In the topological regimes ($\cC\neq 0$), there is an obstruction to construct drives with arbitrarily low dissipation (for a fixed $\tau_2$). 
 The geometric bound $\mathcal{B}_{\rm g}$ tracks the actual dissipation even near the topological phase transition points (at $m=\pm 0.5,\pm 1.5$). Notice that 
 the topological bound $\mathcal{B}_{\rm t}$ is not tight due to large variations of $\gamma$ as a function of $\phi_1$ and $\phi_2$. For Ohmic baths, {which occur in many realistic settings,
we expect $\gamma$ to be approximately constant~\cite{Breuer_Petruccione}.
We therefore also consider a model with a constant value of $\gamma$ by fixing $\Delta(t) = \Delta_0$, leading to a} markedly improved bound $\mathcal{B}_{\rm t}$ 
compared to the time-dependent gap problem, see Fig.~\ref{fig:bounds}c.

Finally, we test the case $\overline g_{12}
\neq 0$. To this end, we fix $\theta=0.2\pi$, $m=1.2$, and vary $\varphi$. The value and sign of $\overline g_{12}$ evolve as a function of $\varphi$, see Fig.~\ref{fig:bounds}e, {while the frequency conversion rate in Fig.~\ref{fig:bounds}f shows that the system correspondingly switches between different topological phases.}
Fig.~\ref{fig:bounds}d compares the dissipation to the topological bound $\mathcal{B}_{\rm t}$ in Eq.~\eqref{eq:BoundSymmetric}. While formally, $\mathcal{B}_{\rm t}$ is not valid when $\overline g_{12}<0$, it provides a natural scale for the dissipation rate which is not reached in this model.

To estimate dissipation rates in this driven-spin example, let us consider an isolated cloud of $N_{\rm c}$ spin-1/2 atoms
subjected to a time-dependent magnetic field~\cite{Zhang2017,Monroe2021,Periwal2021,Feng2023}. Each atom weakly interacts with other identical atoms in its surroundings, acting as an effective heat bath. The energy 
dissipated by $N_{\rm c}$ atoms leads to an increase in the cloud's temperature $T_{\rm c}$,
with the rate $\frac{dT_{\rm c}}{dt}=N_{\rm c} \overline{P}_{\rm d}/C_{\rm c}$, where $C_{\rm c}$ is the heat capacity of the cloud and $\overline{P}_{\rm d}$ the heating rate per spin. 
Assuming a magnetic field amplitude
$|\bB|\sim 2\pi\times 10 \rm \, KHz$, frequencies $\omega_1,\omega_2\sim 2\pi\times 1 \, \rm {KHz}$, and relaxation time $\tau_2\sim 1\,\rm ms$,
we estimate $\gamma\approx 0.03$ and $\overline P_{\rm d}\approx 2\times 10^{-29}\, \rm {Watt}$, using Eq.~\eqref{eq:BoundSymmetric}. Finally, approximating $C_{\rm c}\approx N_{\rm c} k_{\rm B}$, we find $\frac{dT_{\rm c}}{dt}\approx 1.4\, \rm {\mu K/s}$. Given typical cloud temperatures on the order of $T_{\rm c} \sim {\rm nK}$, this heating rate should be experimentally accessible.

\paragraph{Summary and  outlook.---}
In this work, we investigated energy dissipation from a slowly-driven quantum system to a cold heat bath, and showed that the dissipation rate is controlled by 
the quantum geometry of the driving protocol and the quality factor of the system-bath coupling [see Eqs.~\eqref{eq:Wd_omegas} and \eqref{eq:a_tale_of_two_metrics}]. The implications of our findings extend to a range of experimental platforms,
such as nuclear spins~\cite{Jones2000}, solids~\cite{Nathan2022}, superconducting qubits~\cite{Malz2021,Gm2023}, and cavity QED~\cite{Long2022}. 

Remarkably, in a class of driving protocols (such as those hosting topological frequency conversion), the system inevitably generates heat at a rate no lower than that given by Eqs.~\eqref{eq:geometric_bound} and \eqref{eq:BoundSymmetric}, for a symmetric configuration of the drives. 
Interestingly,
for an asymmetric drive configuration, the theoretical minimal dissipation rate can (in principle) be reduced due to interference between the two drives. However, we were not able to identify configurations that violate
the topological bound; this may however provide a promising direction for designing low-dissipation driving protocols.

In this study, we focused on a specific class of isotropic dissipators valid to leading order in diabatic corrections. In general, heat baths derived from microscopic models~\cite{Breuer_Petruccione,Nathan2020,Mozgunov2020} may contain non-isotropic as well as subleading diabatic corrections that may lead to relevant contributions to dissipation; their examination is left for future work.
Another intriguing direction for further study
concerns dissipation in multi-qubit adiabatic computational protocols~\cite{Sarandy2005,Albash2018}, where heating can result in an increased error rate~\cite{Albash2015} and loss of quantum information. Systems with spatial structure (such as quantum pumps~\cite{Citro2023}) also constitute an intriguing platform where driving can induce heat currents. Finally, it would be interesting to explore extensions of our results beyond the average properties of dissipation, considering {\it e.g.} the statistics (or higher moments) of dissipated heat~\cite{Jarzynski2004, Funo2018}.
 
\begin{acknowledgments}

\paragraph{Acknowledgments.---}
We thank Jason Alicea, Israel Klich, Nandagopal Manoj, Johannes Mitscherling, Valerio Peri, and Mark Rudner for insightful discussions. \'E. L.-H. was supported by the Gordon and Betty Moore
Foundation’s EPiQS Initiative, Grant GBMF8682.
F.N. was supported by  the U.S. Department of Energy, Office of Science, Basic Energy Sciences under award DE-SC0019166, the Simons Foundation under award 623768, and the Carlsberg Foundation, grant CF22-0727.
G.R. and I.E. are grateful for support from the Simons
Foundation and the Institute of Quantum Information
and Matter. 
G.R. is grateful for support from the ARO MURI grant FA9550-22-1-0339. This work was performed in
part at Aspen Center for Physics, which is supported by National Science
Foundation grant PHY-221045.
\end{acknowledgments}


%

\end{document}


\title{Quantum geometry and bounds on dissipation in slowly driven quantum systems - Supplementary material
}

\author{Iliya Esin}
\affiliation{Department of Physics and Institute for Quantum Information and Matter, California Institute of Technology, Pasadena, California 91125, USA}
\address{Department of Physics, Bar-Ilan University, 52900, Ramat Gan, Israel}

\author{\'Etienne Lantagne-Hurtubise}
\affiliation{Department of Physics and Institute for Quantum Information and Matter, California Institute of Technology, Pasadena, California 91125, USA}
\affiliation{Département de Physique and Institut Quantique, Université de Sherbrooke, Sherbrooke, Québec, Canada J1K 2R1}

\author{Frederik Nathan}
\affiliation{Department of Physics and Institute for Quantum Information and Matter, California Institute of Technology, Pasadena, California 91125, USA}
\affiliation{Center for Quantum Devices and NNF Quantum Computing Programme, Niels Bohr Institute, University of Copenhagen, 2100 Copenhagen, Denmark}

\author{Gil Refael}
\affiliation{Department of Physics and Institute for Quantum Information and Matter, California Institute of Technology, Pasadena, California 91125, USA}
\date{\today}

\begin{abstract}
Here, we present expanded derivations of the key results in the main text and additional discussions. We assume throughout $\hbar=1$.
\end{abstract}

\maketitle

\begin{figure}[t]
  \centering
  \includegraphics[width=8.6cm]{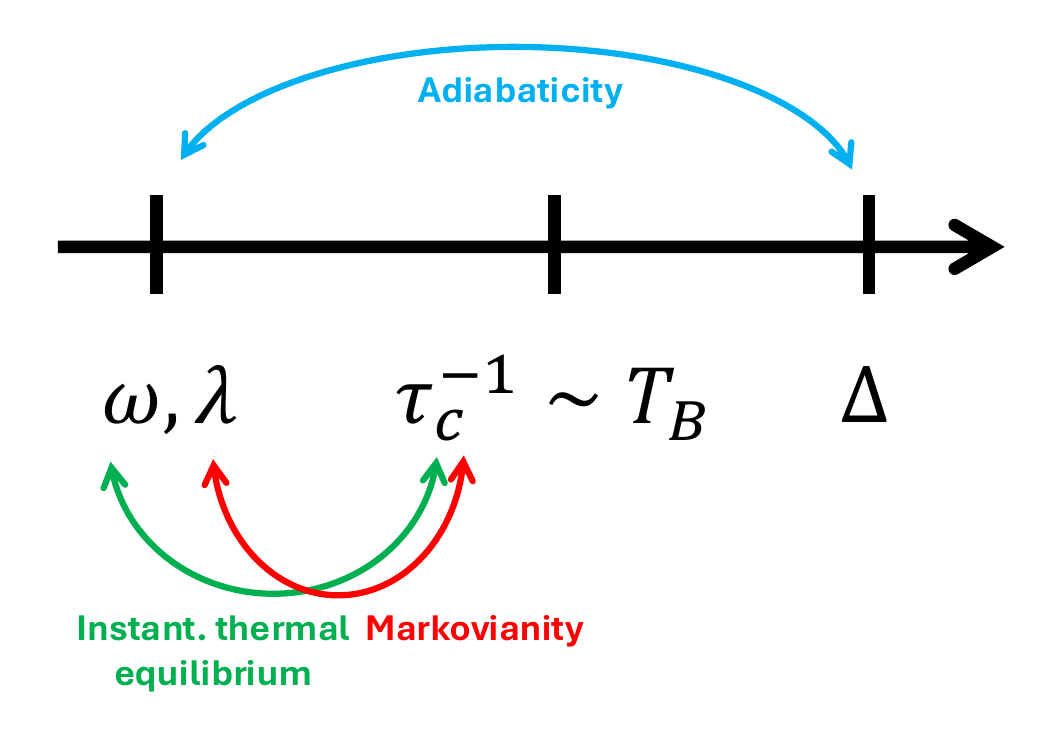}
  \caption{{Hierarchy of energy scales assumed in our work. Here $\omega$ denotes the driving frequency, $\lambda \sim |\!|\Gamma|\!|$ the characteristic system-bath coupling strength, $\tau_c$ the intrinsic correlation time of the bath, $T_B$ the bath temperature and $\Delta$ the energy gap in the system. Various physical conditions imposed on our steady-state solution are highlighted by colored arrows.}}
   \label{fig:OrderScales}
\end{figure}

\section{Hierarchy of energy scales}

To organize our expansion we consider the following hierarchy of energy scales, summarized in Fig.~\ref{fig:OrderScales}. We assume that the gap $\Delta$ is the largest energy scale in the problem, with the driving frequency $\omega$, the system-bath coupling strength $\lambda \sim |\!|\Gamma|\!|$,
and the inverse correlation time of the bath $\tau_c^{-1}$ (commonly determined by the bath temperature $T_{\rm B}$~\cite{Breuer_Petruccione}) all much smaller than $\Delta$. In particular, this hierarchy includes the assumption of adiabaticity, $\omega \ll \Delta$.

In order to neglect non-Markovian corrections arising from the coupling to the bath we further assume that $\lambda \ll \tau_c^{-1} \sim T_{\rm B}$. The assumption of instantaneous thermal equilibrium, whereby the dissipator (at any given time $t$) tends to relax the system towards its instantaneous thermal equilibrium state with the bath, is justified in the limit $\omega \ll \tau_c^{-1}$. However, the relationship between $\omega$ and $\lambda$ can remain general, ranging from the under-damped ($\lambda \ll \omega$) through the over-damped ($\omega \ll \lambda$) regime.

\section{Steady state solution of the Bloch equation}

Here we find the steady state solution to
the Bloch equation [Eq.~(3) in the main text], 
\begin{equation}
    \dot {\vec S}(t) =2 \vd (t)\times  \vS(t) - \Gamma (t)(\vS(t)-\vS_0(t)),
    \label{eq:simple_be}
\end{equation}
and compute the steady-state dissipation rate. In the following we will use the relaxation-time approximation
dissipator that takes the form
\begin{equation}
\Gamma_{ij}(t) = \frac{1}{\tau_1(t)} \hat \bd_i(t)\hat \bd_j(t) + \frac{1}{\tau_2(t)} (\delta_{ij}-\hat \bd_i (t)\hat \bd_j(t)) .
\label{eq:Gamma_app}
\end{equation}
This expression describes the most general isotropic dissipator, i.e., symmetric under rotations around the axis of the instantaneous Hamiltonian represented by $\hat{\bd}(t)$ -- see Eq.~(4) in the main text,  where the 
term proportional to $\delta$ can be absorbed in a redefinition of the instantaneous gap.

We now perform a time-dependent rotation $R(t)$ that maps Eq.~\eqref{eq:simple_be} into an equation of the same form, but in a different frame, where 
$\hat\bd(t)$ points in the $\hat{\bz}$ direction:
\begin{equation}
  R \hat\bd = \hat{\vec z}.
\end{equation}

The spin in the transformed frame, ${\vS}'= R \vS$,
evolves according to the differential equation
\begin{equation}
    \dot {\vS}' = \Delta \hat{\bz} \times \bS'  + \dot R R^{-1} \bS' - \Gamma' (\bS' - \bS'_0) 
\end{equation}
with $\Gamma' = R \Gamma R^{-1}$, and $\bS_0' = R\bS_0$. {At low temperature the system's ground state is parametrized by}
$\bS_0=-\hat{\bd}$,
hence we have $\bS_0' =  -\hat{\vec z}$. The advantage of this frame transformation is that the time dependence of $\bd(t)$ 
has been partially accounted for: $\bd$ now points along the $\hat\bz$ axis and its time dependence only comes through its magnitude $\Delta(t)$. Such a time-dependent frame transformation, however, comes with the Berry connection term $\dot R R^{-1}$ which can be simplified in the following way. First, note that $\dot R R^{-1}$ is a generator of rotations, and thus is a real and anti-symmetric three dimensional matrix -- this can be seen from
\begin{equation}
    \left( \dot R R^{-1} \right)^T = R \dot R^{-1} = - \dot R R^{-1} ,
\end{equation}
where we used $R^{-1} = R^T$ for rotation matrices and $\partial_t ( R R^{-1} ) = 0$. Therefore, the action of such a matrix on the spin vector $\bS'$ can be equivalently represented as a cross product with some real vector $
\bg'$, $\dot R R^{-1} \bS' = \bg' \times \bS'$. In order to find $\bg'$, we notice that because the new frame of reference has a time-independent axis, we can write
\begin{equation}
    \partial_t ( \hat\bz = R \hat{\bd} )  = \dot{R} \hat{\bd} + R \dot{\hat{\bd}} = 0.
    \label{eq:CondR}
\end{equation}
This condition is fulfilled if  $ R$ is given by the solution to the differential equation
\begin{equation}
    \dot R \bv = - R \left[ ( \hat{\bd} \times \dot{\hat{\bd}}) \times \bv \right]
    \label{eq:ExplicitdotR}
\end{equation} 
 for any vector $\bv$, with   initial condition $R(0)=R_0$ where $R_0$ is some orthogonal matrix satisfying $R_0\hat{\bd}(0)=\hat\bz$; note that this differential equation satisfies $\dot R \vec v \perp R\vec v$, implying that the solution is an orthogonal matrix, as we require.
To see why this choice of $ R$ satisfies Eq.~\eqref{eq:CondR}, note that setting $\bv = \hat{\bd}$ results in $\dot{R} \hat{\bd} = - R \dot{\hat{\bd}}$.

Using Eq.~\eqref{eq:ExplicitdotR}, we arrive at
\begin{equation}
\dot R \left( R^{-1} \bS' \right) = - R \left[ ( \hat{\bd} \times \dot{\hat{\bd}}) \times R^{-1} \bS' \right] = - \left[ R ( \hat{\bd} \times \dot{\hat{\bd}})\right] \times \bS' .
\end{equation}
We therefore identify $\bg' = - R (\hat{\bd} \times \dot{\hat{\bd}})$
and write Eq.~\eqref{eq:simple_be} in the transformed coordinate system as 
\begin{equation}
    \dot\bS' = \left[\Delta \hat{\bz} - R (\hat{\bd} \times \dot{\hat{\bd}} ) \right] \times \bS' - \Gamma'(\bS' +
    \hat{\bz} ).
    \label{eq:Bloch_equation_primeframe1}
\end{equation}
where we used $\bS_0'= - \hat\bz$ at zero temperature.

\subsection{Exactly solvable toy model}

To gain intuition, we first consider a solvable toy model,
where the Hamiltonian rotates periodically in time along a circular trajectory with $\bd = \frac12\Delta_0(\sin \omega t, \cos \omega t, 0)$, with a constant gap $\Delta=\Delta_0$. The time-dependent frame rotation $R$ rotates $\hat{\bd}$ to the $\hat{\bz}$ axis; we take the explicit form $R = R_x(\pi/2) R_z(\omega t) $, such that  $R \left( \hat{\bd} \times \dot{\hat \bd} \right) =  \omega \hat{\by}$. The orthonormal rotating-frame basis $(\hat{\bx}, \hat{\by}, \hat{\bz})$ is completed by identifying $R\left( \dot{\hat{\bd}} \right) = \omega \hat{\bx}$, thus preserving the right-hand rule. The Bloch equation in the rotating frame then takes the form

\begin{equation}
   \dot \bS' = \left[\Delta_0 \hat{\bz} - \omega \hat{\by}  \right] \times \bS' - \Gamma'(\bS' + \hat{\bz} ),
   \label{eq:Bloch_equation_primeframeA}
\end{equation}
which is now time independent. The dissipator in the rotating frame is taken to follow the relaxation-time approximation, Eq.~\eqref{eq:Gamma_app}, with $\Gamma' = R \Gamma R^T = {\rm diag} (1/\tau_2, 1/\tau_2, 1/\tau_1)$. The steady-state solution is obtained by setting the left-hand side of Eq.~\eqref{eq:Bloch_equation_primeframeA} to $0$, leading to
\begin{equation}
\bS'_{\rm st} = - \frac{1}{\tau_1} \left( M \right)^{-1}  \hat{\bz} ,
\label{eq:steady_state_primeframe}
\end{equation}
where  $M$ is the matrix defined by $M \vec v = \left[\omega \hat{\by}  - \Delta_0 \hat{\bz} \right] \times\vec v + \Gamma'\vec v$. 
In the example we consider,
\begin{equation}
    M = \begin{pmatrix}
    1/\tau_2 & \Delta_0 & \omega \\
    -\Delta_0 & 1/\tau_2 & 0 \\
    -\omega & 0 & 1/\tau_1 
    \end{pmatrix} 
\end{equation}
in the orthonormal basis defined by $(\hat{\bx}, \hat{\by}, \hat{\bz})$. Inverting $M$ and substituting in Eq.~\eqref{eq:steady_state_primeframe}, we find the steady-state solution 
\begin{equation}
    \bS'_{\rm st} = \frac{1}{1 + \Delta_0^2 \tau_2^2 + \tau_1 \tau_2 \omega^2 }
    \begin{pmatrix}
     \tau_2 \omega \\
     \Delta_0 \omega \tau_2^2 \\
     - 1 - \Delta_0^2 \tau_2^2 
    \end{pmatrix}.
\end{equation}
The dissipated power is obtained through $P_{\rm d} = \dot \bd \cdot \bS_{\rm st} = {\frac{1}{2}} \Delta_0 \dot{\hat{\bd}} \cdot \bS_{\rm st} =  \frac{1}{2} \Delta_0 \omega \hat{ \bx } \cdot \bS'_{\rm st} $,as the transformation between frames maps $(\hat{\bx}, \hat{\by}, \hat{\bz}) \leftrightarrow ( \frac{\dot{\hat{\bd}}}{\omega}, \frac{\hat{\bd} \times \dot{\hat{\bd}}}{\omega}, \hat{\bd})$.
Thus,
\begin{equation}
    P_{\rm d} = \frac{1}{2} \frac{\Delta_0 \tau_2 \omega^2}{1 + \Delta_0^2 \tau_2^2 +  \tau_1 \tau_2 \omega^2 } .
\end{equation}

\subsection{Back to the general case}

With the intuition from the above toy model in mind, we now come back to the case where the system is driven by an arbitrary (but slow) drive. The problem posed by Eq.~\eqref{eq:Bloch_equation_primeframe1} is now time dependent, but the time dependence is of order $\omega/\Delta$, 
where $\omega\sim |\dot {\hat \bd}|$. More precisely, as we will see we can once again set the left-hand-side of Eq.~\eqref{eq:Bloch_equation_primeframe1} to zero, because $\dot \bS'$ only contains terms at order $\cO ( \omega^2/\Delta^2)$ or higher. 
 
{We can then solve for the (now time-dependent) steady state in the rotating frame in a similar way. To parametrize a more general time evolution we consider 
\begin{equation}
    R \left( \hat{\bd} \times \dot{\hat{\bd}} \right) = | \dot{\hat{\bd}}| R(\hat{\vec f}) = | \dot{\hat{\bd}}|  \left(a \hat{\bx} + b \hat{\by} \right),
    \label{eq:app_general_rotation}
\end{equation}
with $a$ and $b$ time-dependent functions parametrizing rotations, normalized such that at all times $a^2 + b^2 = 1$, and we defined the unit vector $\hat{\vec f} \equiv  \left( \hat{\bd} \times \dot{\hat{\bd}} \right) / |\hat{\bd} \times \dot{\hat{\bd}}| = \left( \hat{\bd} \times \dot{\hat{\bd}} \right) / | \dot{\hat{\bd}}|$. (Note that a component proportional to $\hat\bz$ is not allowed in Eq.~\eqref{eq:app_general_rotation} because  $\hat{\bd} \times \dot{\hat{\bd}} $ must always be perpendicular to $\hat{\bd}$.) To complete the basis transformation, we define the unit vector $\hat{\vec e} = \dot{\hat{\bd}}/|\dot{\hat{\bd}}|$ such that $\hat{\vec f}  = \hat{\vec d}  \times \hat{\vec e}$. In the rotating frame basis this unit vector is mapped to
\begin{equation}
    R \left( \hat{\be} \right) = \left(b \hat{\bx} - a \hat{\by} \right) .
\end{equation}
}

{The instantaneous steady-state equation {in the rotating $(\hat{\bx}, \hat{\by}, \hat{\bz})$} basis follows the form of Eq.~\eqref{eq:steady_state_primeframe}, but with a time-dependent matrix
\begin{equation}
    M = \begin{pmatrix}
    1/\tau_2 & \Delta & b | \dot{\hat{\bd}}| \\
    -\Delta & 1/\tau_2 & -a | \dot{\hat{\bd}}| \\
    -b | \dot{\hat{\bd}}| & a | \dot{\hat{\bd}}| & 1/\tau_1 
    \end{pmatrix} .
    \label{eq:M_matrix}
\end{equation}
}
Its expression is simpler when transforming back to the original (laboratory) frame with basis vectors $(\hat{\be}, \hat{\vec f}, \hat{\bd})$:
{
\begin{equation}
    \bS_{\rm st} = \frac{1}{1 + \Delta^2 \tau_2^2 +  \tau_1 \tau_2 |\dot{\hat{\bd}}|^2}
    \begin{pmatrix}
    { |\dot{\hat{\bd}}| \tau_2 } \\
    |\dot{\hat{\bd}}| \Delta \tau_2^2 \\
    -1- \Delta^2 \tau_2^2 
    \end{pmatrix} ,
    \label{eq:S_st_labframe}
\end{equation}
as reported in the main text, see Eq.~(5).}
This steady-state solution admits a time derivative that is at least of order $\omega^2/\Delta^2$, thus justifying our setting the left-hand side of Eq.~\eqref{eq:Bloch_equation_primeframe1} to zero to order $\omega/\Delta$. 

We then compute dissipation, using the laboratory-frame steady state obtained in Eq.~\eqref{eq:S_st_labframe}:
\begin{align}
    P_{\rm d} = \dot \bd \cdot \bS_{\rm st} =&  \frac1 2\Delta \dot{\hat{\bd}} \cdot \bS_{\rm st} + \frac 1 2\dot \Delta \hat{\bd} \cdot \bS_{\rm st} \nonumber \\
     =& \frac{1}{2} \frac{\Delta \tau_2 |\dot{\hat{\bd}}|^2 - \dot{\Delta} ( 1 + \Delta^2 \tau_2^2)}{1 + \Delta^2 \tau_2^2 + \tau_1 \tau_2 |\dot{\hat{\bd}}|^2 }.
\end{align}
The first term leads to geometric dissipation (see Eq.~(6) in the main text). The second term averages to zero over a period when the diabatic correction $\sim |\dot{\hat{\bd}}|^2$ in the denominator can be neglected, as it can then be re-expressed as a total time derivative.

\subsection{Finite-temperature extension}

We now consider contributions to dissipation when the heat bath is at a non-zero temperature $T_B$. Assuming an isotropic dissipator, we have $\bS_0 = -\tanh(\beta \Delta/2) \hat{\bd}$ which points against the direction of the instantaneous Hamiltonian. Therefore, we simply need to solve a generalization of the Bloch equation in the rotating-frame basis, Eq.~\eqref{eq:Bloch_equation_primeframe1}, with $\bS_0' = - \tanh(\beta \Delta/2) \hat{\bz}$:
\begin{equation}
   \dot\bS' = \left[\Delta \hat{\bz} - R (\hat{\bd} \times \dot{\hat{\bd}} ) \right] \times \bS' - \Gamma'\left(\bS' +
   \tanh\frac{\beta \Delta}{2} \hat{\bz}\right).
   \label{eq:Bloch_equation_primeframeT}
\end{equation}

Following the steps outlined above, we are interested in finding a solution to the above equation which is stationary to linear order in $\omega$. 
One could naively attempt to generalize the arguments applied to the low-temperature limit, leading to $\bS'_{\rm st} = \vec v$, where $\vec v \equiv -\frac{M^{-1}}{\tau_1}\tanh(\beta \Delta/2)\hat{\bz}$, with the $M$ matrix given by Eq.~\eqref{eq:M_matrix}. 
Explicitly, 
\begin{equation}
    \vec v = \frac{\tanh(\beta \Delta/2)}{1  + \Delta^2 \tau_2^2 + \tau_1 \tau_2 |\dot{\hat{\bd}}|^2 }
    \begin{pmatrix}
        |\dot{\hat{\bd}}| \tau_2 \\ 
        |\dot{\hat{\bd}}| \Delta \tau_2^2 \\
         -1 - \Delta^2 \tau_2^2 
    \end{pmatrix} .
    \label{eq:S_st_labframe_finiteT}
\end{equation}
However, this solution does not constitute a legitimate
solution to linear order in $\omega/\Delta$, because its time derivative $\dot{ \bS}_{\rm st}$ contains a  term linear in $\omega$:
\begin{equation}
    \dot{\vec v} = -\frac{\beta \dot \Delta/2}{\cosh^2(\beta \Delta/2)} \hat{\bz} + {\cal O} \left( \frac{\omega^2}{\Delta} \right).
    \label{eq:S_st_timederivative}
\end{equation}
We thus need to perform one more step: solving Eq.~\eqref{eq:Bloch_equation_primeframeT} with the left-hand side replaced by Eq.~\eqref{eq:S_st_timederivative} instead of set to zero. Doing this, we obtain the steady-state
\begin{equation}
\tilde{\bS}'_{\rm st} = -  M ^{-1} \left( \frac{\tanh(\beta \Delta/2)}{\tau_1} - \frac{\beta \dot \Delta/2}{\cosh^2(\beta \Delta/2)} \right) \hat{\bz} ,
\end{equation}
Explicitly,
\begin{equation}
    \tilde{\bS}'_{\rm st} = \frac{\left(\tanh(\beta \Delta/2) - \tau_1\frac{\beta \dot \Delta/2}{\cosh^2(\beta \Delta/2)} \right)}{1 + \Delta^2 \tau_2^2 + \tau_1 \tau_2 |\dot{\hat{\bd}}|^2 }
    \begin{pmatrix}
        |\dot{\hat{\bd}}| \tau_2 \\
        |\dot{\hat{\bd}}| \Delta \tau_2^2 \\
        -1 - \Delta^2 \tau_2^2 
    \end{pmatrix} .
    \label{eq:finite_T_steadystate}
\end{equation}
One can check that the Bloch equation Eq.~\eqref{eq:Bloch_equation_primeframeT} is satisfied to linear-order in $\omega$ by inserting $\tilde{\bS}'_{\rm st}$ in Eq.~\eqref{eq:finite_T_steadystate}, thus providing a legitimate diabatic expansion.

Having found the steady-state solution in Eq.~\eqref{eq:finite_T_steadystate}, we next compute the dissipated power, {obtaining
\begin{widetext}
\begin{align}
    P_{\rm d} =&  \frac{1}{2} \Delta \dot{\hat{\bd}} \cdot \bS_{\rm st} + \frac 1 2\dot \Delta \hat{\bd} \cdot \bS_{\rm st} 
     = \frac{\Delta |\dot{\hat \bd}|}{2}\hat{\bx}\cdot \bS_{\rm st}' +\frac{\dot\Delta}{2}\hat{\bz}\cdot \bS_{\rm st}'\\ 
     =& \frac{ \tanh \frac{\beta \Delta}{2} \left( \Delta \tau_2 |\dot{\hat{\bd}}|^2  - \dot \Delta \left(1 + \Delta^2 \tau_2^2 \right) \right) + \frac{\tau_1 \beta}{2 \cosh^2\frac{\beta \Delta}{2}} \left( - \Delta \dot \Delta \tau_2  |\dot{\hat{\bd}}|^2 + \dot \Delta^2 \left(1 + \Delta^2 \tau_2^2 \right) \right)}{2 \left(1 + \Delta^2 \tau_2^2 +  \tau_1 \tau_2 |\dot{\hat{\bd}}|^2 \right)}.
\end{align}
\end{widetext}
First of all, we neglect the term that contains three time derivatives, as it is subleading in the adiabatic limit (where $\omega, 1/\tau_2 \ll \Delta$, see Fig.~\ref{fig:OrderScales}). Furthermore,
if $\tau_1$ and $\tau_2$ are of the same order, we can also neglect the term $|\dot{\hat{\bd}}|^2$ term in the denominator. In this case, as also explained in the main text, the term proportional to $\dot \Delta$ vanishes, as it can be rewritten as a total time derivative. We then arrive at a much simpler expression for finite-temperature dissipation,
\begin{align}
    P_{\rm d}
     =& \frac{1}{2} \frac{\Delta \tau_2 \tanh \frac{\beta \Delta}{2} }{1 + \Delta^2 \tau_2^2} |\dot{\hat{\bd}}|^2 + \frac{\tau_1 \beta}{4 \cosh^{2}\frac{\beta \Delta}{2}} \dot{\Delta}^2.
\end{align}

The first term above is simply the finite-temperature generalization of the quantum geometric dissipation discussed in the main text. The second term is intrinsically a finite-temperature dissipation effect that arises from the time-dependence of the gap magnitude $\Delta$. This term depends linearly on the longitudinal relaxation time scale $\tau_1$, and is suppressed exponentially as $\beta e^{-\beta \Delta}$ at low temperatures. 

\section{Generic lower and upper bounds on the dissipation}

Here, we derive the lower bound on the dissipation away from the symmetric case 
and an upper bound. As follows from Eq.~(11) in the main text, the average dissipation is given by 
\begin{equation}
\overline P_{\rm d}=\frac{1}{T}\int_0^T dt\gamma [\omega_1^2 g_{11}+\omega_2^2 g_{22}+2\omega_1\omega_2 g_{12}].
\label{eq:DissipationDefinition}
\end{equation}
For convenience, we parametrize, $\sqrt{g_{11} g_{22}}=r$, and $g_{12}=r\cos(\theta)$, using two parameters, $\theta=[-\pi,\pi]$ and $r$. From the definition of the Berry curvature, it follows $\Omega=2r \sin(\theta)$. Next we use the inequality $2r\le g_{11}+g_{22}$, and $2r\ge\Omega$, to find  
\begin{equation}
    \frac{|\cC|}{4\pi}\le\oiint \frac{d\phi_1d\phi_2}{4\pi^2} r\le \frac{\cP_1+\cP_2}{2\Delta_{\rm min}^2},
    \label{eq:InequalityR}
\end{equation}
where $\cP_i=\oiint \frac{d\phi_1d\phi_2}{4\pi^2} (\partial_i \bd)^2$. 

\subsection{Lower bound}
We begin with the lower bound, in this case we can use the inequality 
$\omega_1^2 g_{11}+\omega_2^2 g_{22}+2\omega_1\omega_2 g_{12}\ge 2\omega_1\omega_2 r[1+\cos(\theta)]\ge \omega_1\omega_2 r[1-\cos^2(\theta)]=\omega_1\omega_2 r \sin^2(\theta)$. 
Therefore, using Eq.~\eqref{eq:DissipationDefinition}, the bound on the dissipation reads 
\begin{equation}
    \overline{P}_{\rm d}\ge \omega_1\omega_2\oiint \frac{d\phi_1d\phi_2}{4\pi^2}\gamma r\sin^2(\theta).
\end{equation}
Next, we rewrite $\theta$ in terms of $\Omega$, using the parametrization of $\Omega$, to arrive at
\begin{equation}
    \overline{P}_{\rm d}\ge \omega_1\omega_2\oiint \frac{d\phi_1d\phi_2}{4\pi^2}\gamma \frac{\Omega^2}{4r}.
    \label{eq:DissipationBound}
\end{equation}
To make progress, we use the following sequence of inequalities, using the Cauchy–Schwarz inequality, to obtain
\begin{widetext}
\begin{equation}
    \frac{|\cC|}{2\pi} = |\oiint \frac{d\phi_1d\phi_2}{4\pi^2}\Omega|\le \oiint \frac{d\phi_1d\phi_2}{4\pi^2}|\Omega|=\oiint \frac{d\phi_1d\phi_2}{4\pi^2}\frac{|\Omega| \sqrt{r}}{\sqrt{r}}\le\sqrt{\oiint \frac{d\phi_1d\phi_2}{4\pi^2}\frac{\Omega^2}{r}\oiint \frac{d\phi_1d\phi_2}{4\pi^2}r} .
\end{equation}
\end{widetext}
Rearranging and squaring the first and last terms of the inequality gives rise to
\begin{equation}
    \oiint \frac{d\phi_1d\phi_2}{4\pi^2}\frac{\Omega^2}{r}\ge \frac{\cC^2}{4\pi^2\oiint \frac{d\phi_1d\phi_2}{4\pi^2}r}.
\end{equation}
Next, using the inequality in Eq.~\eqref{eq:InequalityR}, we obtain
\begin{equation}
    \oiint \frac{d\phi_1d\phi_2}{4\pi^2}\frac{\Omega^2}{r}\ge \frac{\Delta_{\rm min}^2\cC^2}{2\pi^2(\cP_1+\cP_2)}.
    \label{eq:Ineq1}
\end{equation}

Finally, using $\gamma\ge\gamma_{\rm min}$ in Eq.~\eqref{eq:DissipationBound}, taking $\gamma_{\rm min}$ outside the integral and using Eq.~\eqref{eq:Ineq1}, we arrive at the lower bound on the dissipation,
\begin{equation}
    \overline{P}_{\rm d}\ge \frac{\omega_1\omega_2\gamma_{\rm min}\Delta_{\rm min}^2\cC^2}{8\pi^2(\cP_1+\cP_2)}.
\end{equation}

\subsection{Upper bound}

Next, we proceed to find the upper bound on $\overline P_{\rm d}$. 
Here, we use the inequality 
$g_{11}\omega_1^2+g_{22}\omega_2^2+2g_{12}\omega_1\omega_2\le2(g_{11}\omega_1^2+g_{22}\omega_2^2)$.

Next, we use $\gamma\le\gamma_{\rm max}$, and take take $\gamma_{\rm max}$ outside of the integral, to arrive at
\begin{equation}
    \overline{P}_{\rm d}\le 2\gamma_{\rm max}\oiint \frac{d\phi_1d\phi_2}{4\pi^2} [g_{11}\omega_1^2+g_{22}\omega_2^2].
\end{equation}

Next, we use an explicit expression $g_{11}=\frac{1}{\Delta^2}[(\partial_1\bd)^2-(\partial_1\Delta)^2]$, to obtain $g_{11}\le\frac{(\partial_1\bd)^2}{\Delta_{\rm min}^2}$, and similarly for $g_{22}$, yielding the uppper bound on the dissipation,
\begin{equation}
    \overline{P}_{\rm d}\le 2\gamma_{\rm max}\frac{\omega_1^2\cP_1+\omega_2^2\cP_2}{\Delta_{\rm min}^2}.
\end{equation}
 
\section{Symmetry conditions for vanishing of the averaged $g_{12}$}

Here, we find the conditions on the symmetries of the system for $\int \frac{d\phi_1 d\phi_2}{4\pi^2}g_{12}=0$.
We consider a two-level system irradiated by two monochromatic drives, coupled to the spin trough 
\begin{equation}
\bd = \bB_1(\phi_1) + \bB_2(\phi_2) + \vec m,
\label{eq:d_Vector}
\end{equation}
where $\phi_i=\omega_i t$. We use the following definition of the quantum metric, which can be derived from $g_{ab}=\frac{1}{4}\partial_{\phi_a} \hat\bd\cdot\partial_{\phi_b}\hat \bd$, reading 
\begin{equation}
    g_{ab} = \frac{3}{16}\frac{\partial_{\phi_a}\partial_{\phi_b}d^2}{d^2}-\frac{1}{4}\frac{\bd\cdot\partial_{\phi_a}\partial_{\phi_b}\bd}{d^2}-\frac{1}{32}\frac{\partial_{\phi_a}\partial_{\phi_b}d^4}{d^4},
\end{equation}
where $d=|\bd|$. For the form of $\bd$ in Eq.~\eqref{eq:d_Vector}, the second term in the RHS of $g_{12}$ vanishes, $\partial_{\phi_1}\partial_{\phi_2}\vec d=0$, giving rise to 
\begin{equation}
    g_{12} = \frac{3}{16}\frac{\partial_{\phi_1}\partial_{\phi_2}d^2}{d^2}-\frac{1}{32}\frac{\partial_{\phi_1}\partial_{\phi_2}d^4}{d^4},
    \label{eq:g12}
\end{equation}
which is only expressed in terms of the scalar $d^2=\bB_1^2(\phi_1)+\bB_2^2(\phi_2)+\vm^2+2\bB_1(\phi_1)\cdot \bB_2(\phi_2)+2\bB_1(\phi_1)\cdot\vm+2\bB_2(\phi_2)\cdot\vm$.

We are interested in finding the symmetry conditions that enforce that $g_{12}$ in Eq.~\eqref{eq:g12} averages to zero when averaging over the phases $\phi_1$ and $\phi_2$. For this to occur, the norm $d^2$ should be even under reversal of either one of the phases, $\phi_a \rightarrow - \phi_a + c$ with $a=1$ or $2$ and $c$ a constant. Then, the corresponding derivative $\partial_{\phi_a}$ ensures that $g_{12}$ is an odd function over the period of $\phi_a$, and the integral therefore vanishes. Here, we will require $g_{12}$ to be odd in $\phi_1$.

We consider a generic situation where both drives are elliptically polarized with arbitrary orientations in three dimensions,
\begin{align}
    \bB_1(\phi_1) &= b_{1,e}(\phi_1) \hat{\bm e}_1 + b_{1,o}(\phi_1) \hat{\bm e}_2 ,\\
    \bB_2(\phi_2) &= b_{2,e}(\phi_2) \hat{\bm e}_3 + b_{2,o}(\phi_2) \hat{\bm e}_4.
\end{align}
where the principal axes of the two ellipses are respectively orthogonal, $\hat{\bm e}_1 \perp \hat{\bm e}_2$ and $\hat{\bm e}_3 \perp \hat{\bm e}_4$.
Without loss of generality, we choose the origin of time such that the functions $b_{a,e}$ and $b_{a,o}$ are respectively even and odd under $\phi_a \rightarrow -\phi_a$ (in other words, at $\phi_1=0$ the polarization vector lies at along $\hat{\bm e}_1$.). Therefore we have $\bB_a^2(\phi_a) = b_{a,e}^2(\phi_a) + b_{a,o}^2(\phi_a)$ which is even under reversing $\phi_a$.
There are two non-trivial terms that impose conditions on the driving protocol:
\begin{align}
    \bB_1\cdot \bB_2 &= b_{1,e}(\phi_1) b_{2,e}(\phi_2) \hat{\bm e}_1 \cdot \hat{\bm e}_3 +  b_{1,e}(\phi_1) b_{2,o}(\phi_2) \hat{\bm e}_1 \cdot \hat{\bm e}_4 \nonumber \\
    &+ b_{1,o}(\phi_1) b_{2,e}(\phi_2) \hat{\bm e}_2 \cdot \hat{\bm e}_3 +  b_{1,0}(\phi_1) b_{2,0}(\phi_2) \hat{\bm e}_2 \cdot \hat{\bm e}_4
    \label{eq:FdotG}
\end{align}
and
\begin{align}
    \bB_1 \cdot\vm &= b_{1,e}(\phi_1) \hat{\bm e}_1 \cdot \vm + b_{1,o}(\phi_1) \hat{\bm e}_2 \cdot \vm.
    \label{eq:FdotM}
\end{align}
We want both functions to be even under $\phi_1$. This requires that the odd contribution, $b_{1,o}$, vanishes.
To this end, we require $\hat{\bm e}_2 \perp \hat{\bm e}_3, \hat{\bm e}_4$ in Eq.~\eqref{eq:FdotG}. The two drives, therefore, form a right-angle triad of vectors (remember that also $\hat{\bm e}_3 \perp \hat{\bm e}_4$ due to their being the principal axes of an ellipse.) Alternatively, that corresponds to ${\bm S}_1 \perp {\bm S}_2$, where  ${\bm S}_a=\bB_a\times \partial_{\phi_a} \bB_a$ is the angular momentum of the drive.
In addition, Eq.~\eqref{eq:FdotM} leads to a condition on the relative orientation of the mass term and the drive, $\hat{\bm e}_2 \perp \vm$. Because ${\bm S}_2$ is parallel to $\hat{\bm e}_2$ this condition is equivalent to ${\bm S}_2 \perp \vm$.

The argument above can be run using $\phi_2$ instead. The condition on the relative angular momentum vectors of the two drives being perpendicular remains unchanged, whereas the condition that invokes the mass term becomes instead $\hat{\bm S}_1 \perp \vm$. Because only one of the $\phi_a$ integrals needs to vanish, either condition is sufficient. In summary, if ${\bm S}_1 \perp {\bm S}_2$ and \emph{either} ${\bm S}_1 \perp {\bm M}$ or ${\bm S}_2 \perp {\bm M}$, the phase-average of $g_{12}$ will vanish.

\section{Dissipation in a system with an arbitrary number of bands}

Here, we find the energy dissipation rate for an arbitrary number of bands of (non-degenerate) bands, generalizing Eq.~(8) in the main text.
We start from the master equation  which reads (dropping explicit dependence on $t$),
\begin{equation}
    \dot\rho=-i[H,\rho]+\cD\{ \rho\}.
    \label{eq:MasterEquation2}
\end{equation}
Next, we transform Eq.~\eqref{eq:MasterEquation2} to the instantaneous diagonal basis of $H$ by applying a unitary transformation $U$. Such that $U H U^\dagger=H_d$, where $H_d$ is diagonal and $\rho_d = U \rho U^\dagger$, leading to
\begin{equation}
    \dot\rho_d=-i[H_d+\cA^i \dot \alpha^i,\rho_d]+\cD'\{\rho_d\},
    \label{eq:MasterEquation_App}
\end{equation}
where $\cA^i\dot\alpha^i=i\dot UU^\dagger$, denotes the Berry connection tensor $\cA^i_{mn}(\balpha)=-i\bra{\psi_m(\balpha)}\partial_{\alpha^i}\ket{\psi_n(\balpha)}$ for bands labeled by $m,n$, and $\cD'=U\cD U^\dagger$, with $\ket{\psi_m(\balpha)}$ the $m$th eigenstate of $H(\balpha)$. (Here the index $i$ denotes a spatial component.) We consider the full density matrix $\rho_d$, written in the instantaneous eigenbasis of $H$ as 
\begin{equation}
  \rho_{mn} = \rho^{\rm eq}_{m} \delta_{mn} + \delta \rho_{mn}.
  \label{eq:RhoExpansion}
\end{equation}
(Throughout, we suppress the index ``$d$'' for the elements of $\rho$.)
The diagonal contributions $\rho^{\rm eq}_{m}$ for each band denote their corresponding equilibrium values, whereas $\delta\rho_{mn}$ denote the diabatic corrections. The form of the dissipator is chosen as a generalization of the main text expression, Eq.~(4),
\begin{equation}
   \cD'\{\rho\} _{mn} = - \delta \rho_{mn}/\tau_{mn}.
\end{equation}
where  $\tau_{mn}$ denotes the inter-band relaxation timescale associated with transitions between bands $m$ and $n$. This form of the dissipator can be derived from first principles in the limit of slow driving,
provided that the relaxation rate is much smaller than any spectral gap of the system~\cite{DiMeglio_2023}.

We now evaluate the commutator $C = [H_d + \cA^i \dot \alpha^i, \rho ]$ in Eq.~\eqref{eq:MasterEquation_App}. It is helpful to separate $C$ in diagonal and off-diagonal components. The diagonal components read
\begin{align}
    C_{mm} &= \sum_{p \neq m} \left[ \left( H_d +  \cA^i \dot \alpha^i \right)_{mp} \rho_{pm} -  \rho_{mp} \left(H_d +  \cA^i \dot \alpha^i \right)_{pm} \right] \nonumber \\
    &= \sum_{p \neq m} \left[ \cA_{mp}^i \rho_{pm} -  \rho_{mp} \cA_{pm}^i  \right] \dot \alpha^i
\end{align}
where the contributions with $p=m$ vanish trivially. The off-diagonal contributions (with $m \neq n$) read
\begin{align}
    C_{mn} =& \sum_{p} \left[ \left( H_d +  \cA^i \dot \alpha^i \right)_{mp} \rho_{pn} -  \rho_{mp} \left(H_d +  \cA^i \dot \alpha^i \right)_{pn} \right] \nonumber \\
    =& \sum_{p \neq {m,n}} \left[ \cA_{mp}^i \rho_{pn} -  \rho_{mp} \cA_{pn}^i  \right] \dot \alpha^i
    +\cA_{mn}^i \dot \alpha^i \left( \rho_{nn} - \rho_{mm} \right) \nonumber \\ 
    &+ (E_m + \cA_{mm}^i  \dot \alpha^i - E_n -  \cA_{nn}^i  \dot \alpha^i ) \rho_{mn}, \end{align}
where $E_m$ is the energy of the level $m$.
Taking $\dot \rho = 0$ to find steady-state solutions to the master equation leads to a separate condition for each of its components. First consider the off-diagonal ones:
\begin{equation}
 0 = -i C_{mn} - \frac{\delta \rho_{mn}}{\tau_{mn}}
    \label{eq:MasterEquation_App_Solution}
\end{equation}
Solving for the diabatic correction $\delta \rho_{mn}$,
\begin{equation}
    \delta \rho_{mn} = \frac{\cA_{mn}^i \dot \alpha^i \left( \rho^{\rm eq}_{n} - \rho^{\rm eq}_{m} \right) + \sum_{p} \left[ \cA_{mp}^i \delta \rho_{pn} -  \delta \rho_{mp} \cA_{pn}^i  \right] \dot \alpha^i }{E_m + \cA_{mm}^i \dot \alpha^i - E_n -  \cA_{nn}^i \dot \alpha^i - i /\tau_{mn} },
\end{equation}
where the sum is over $p \neq m,n$. 
To leading-order the terms in the sum can be neglected because they contain additional factors of $\dot \alpha^i$ coming from off-diagonal contributions to $\rho$ from other transition to other bands $p$. Similarly, the contribution from the quantum geometric potential in the denominator can be neglected as it produces a contribution quadratic in $\dot \alpha^i$. We thus have
\begin{equation}
    \delta \rho_{mn} = \frac{\cA_{mn}^i \dot \alpha^i \left( \rho^{\rm eq}_{m} - \rho^{\rm eq}_{n} \right)}{E_m - E_n - i /\tau_{mn} }.
    \label{eq:SteadyStateDensityMatrix_app}
\end{equation}
Similarly, the diagonal components read
\begin{align}
    \delta \rho_{mm} =& -i \tau_{mm} \sum_{p \neq m} \left[ \cA_{mp}^i \delta \rho_{pm} -  \delta \rho_{mp} \cA_{pm}^i \right] \dot \alpha^i \nonumber \\
    =& 2 \tau_{mm} \sum_{p \neq m} {\rm Im} \left[ \cA_{mp}^i \delta \rho_{pm}  \right] \dot \alpha^i.
    \label{eq:Correction1}
\end{align}
Notice that we focused on a zero temperature case, in which $\dot\rho^{\rm eq}=0$. It follows from Eq.~\eqref{eq:Correction1} that the diagonal components are of higher order in the diabatic expansion, and thus can be neglected. 

The average dissipation rate is given by $\overline P_{\rm d}=\frac{1}{T}\int_0^T dt{\rm Tr}[\dot H \rho].$
On a diagonal basis, the dissipation reads 
\begin{equation}
    \overline P_{\rm d}=\frac{1}{T}\int_0^T dt{\rm Tr}\{-i[ H_d,\cA^i\dot\alpha^i] \rho_d+\dot H_d\rho_d\}.
    \label{eq:Dissipation1}
\end{equation}
The commutator in the trace  can be explicitly evaluated, resulting in a purely off-diagonal matrix $[H_d,\cA^i \dot \alpha^i]_{mn} = \cA^i_{mn} \dot \alpha^i ( E_m - E_n)$.
To continue, we use the expansion of $\rho_d$ in Eq.~\eqref{eq:RhoExpansion}. The term proportional to $\rho^{\rm eq}$ vanishes under the time average. Therefore, we need to use the second term $\delta \rho_d$ up to linear order in $\dot\balpha$. As we found in Eqs.~\eqref{eq:SteadyStateDensityMatrix_app} and \eqref{eq:Correction1}, this order only appears in the off-diagonal elements of $\delta \rho_d$. Therefore, only the term proportional to a commutator in Eq.~\eqref{eq:Dissipation1} remains, yielding $\overline P_{\rm d}=\frac{1}{T}\int_0^T dt\Lambda^{(N)}_{ij}\dot \alpha^i\dot \alpha^j$ where
\begin{equation}
    \Lambda^{(N)}_{ij}= -i \sum_{m \neq n} \left( \rho^{\rm eq}_{m} - \rho^{\rm eq}_{n} \right)  \frac{\cA^i_{mn}\cA^j_{nm}}{1 + i /\Delta_{mn} \tau_{mn} }.
\end{equation}
Here we used that $\tau_{nm} = \tau_{mn}$ and defined the spectral gap $\Delta_{mn} = E_m - E_n$. Equivalently, we can obtain a convenient expression for the metric by writing the sum as
\begin{widetext}
\begin{align}
    \Lambda^{(N)}_{ij} &= -i \sum_{m < n} \left( \rho^{\rm eq}_{m} - \rho^{\rm eq}_{n} \right) \left\{ \frac{\cA^i_{mn} \cA^j_{nm}}{1 + i /\Delta_{mn} \tau_{mn}} - \frac{\cA^i_{nm} \cA^j_{mn}}{1 - i /\Delta_{mn} \tau_{mn}} \right\}\nonumber \\
    &= 2 \sum_{m < n} \left( \rho^{\rm eq}_{n} - \rho^{\rm eq}_{m} \right) {\rm Im} \left\{ \frac{\cA^i_{mn} \cA^j_{nm}}{1 + i / \Delta_{mn} \tau_{mn}} \right\}.
\end{align}
\end{widetext}
Contrary to the two-level version, the expression inside the bracket cannot be identified with the quantum geometric tensor for band $m$.

{In the low-temperature limit, where only the ground-state energy level is occupied in the instantaneous equilibrium state ($\rho^{\rm eq}_0 = 1$ and $\rho^{\rm eq}_{m\neq0} = 0$),} the above expression simplifies to
\begin{align}
    \Lambda^{(N)}_{ij} 
    &= -2 \sum_{n\neq 0} {\rm Im} \left\{ \frac{\cA^i_{0n} \cA^j_{n0}}{1 + i / \Delta_{0n} \tau_{0n}} \right\}
    \label{eq:metric_steadystate_app} ,
\end{align}
which can be rewritten as 
\begin{equation}
    \Lambda^{(N)}_{ij} 
    = 2 \sum_{n\neq 0}  \frac{\cA^i_{0n} \cA^j_{n0}}{ \Delta_{0n} \tau_{0n} + \Delta_{0n}^{-1} \tau_{0n}^{-1}}.
\end{equation}
This is a positive semidefinite matrix, since for any real vector $\vec v$, 
\begin{equation}
    \vec v^T  \Lambda^{(N)} \vec v 
    = 2 \sum_{n\neq 0}  \frac{|v_{0n}|^2}{ \Delta_{0n} \tau_{0n} + \Delta_{0n}^{-1} \tau_{0n}^{-1}}\geq 0 
\end{equation}
where $v_{0n}=\sum_ i \cA ^i_{0n}v_i$ and we exploited $\cA^ i_{0n} = (\cA^i_{n0})^*$.

We  establish the bound quoted in the main text
using the identity
$ 1 \geq \frac{2}{ \Delta_{0n} \tau_{0n} + \Delta_{0n}^{-1} \tau_{0n}^{-1}}\geq \gamma^{(N)}$, where
\begin{equation}
\gamma^{(N)} = \min_n \frac{2}{\Delta_{0n} \tau_{0n} + \Delta_{0n}^{-1} \tau_{0n}^{-1}} .
\end{equation}
This leads to 
\begin{align}
G^{(N)} \succeq \Lambda^{(N)} \succeq \gamma ^{(N)}G^{(N)},
\end{align}
as quoted in Eq.~(10) of the main text,
where
\begin{equation}
    G^{(N)}_{ij} =  \sum_n \cA^i_{0n}\cA^j_{n0} =\frac{1}{2} {\rm Tr}[\partial_{\alpha^i} P_0 \partial_{\alpha^j} P_0]
\end{equation}
is the ground state quantum metric, and $P_0$ denotes a ground-state projector. Here the inequality $\succeq$ between  metrics $A$ and $B$ should be understood such that $A\succeq B$ if and only if  $\vec v^T A\vec v \geq \vec v^T B\vec v \geq 0$ for all real-valued vectors $\vec v$. 
The resulting inequality translates directly to a corresponding inequality for the energy dissipation.

\section{Derivation of the Master equation from a microscopic model}

Here we derive the master equation appearing in Eq.~(3) in the main text from a microscopic model. We consider a cloud of $N_{\rm c}+1$ spins coupled to a time-dependent magnetic field described by the Hamiltonian
\begin{equation}
H=\sum_{i=0}^{N_{\rm c}} \bd_0(t)\cdot \bsigma_i+\sum_{i>j} J_{ij} \bsigma_i\cdot \bsigma_j.
\end{equation}
We assume a weakly coupled cloud, with $\sum_j |J_{ij}|\ll |\bd_0|$, such that the spins are approximately aligned with $-\bd_0(t)$ at any time, up to small fluctuations. 
Without loss of generality, we focus on a single spin with $i=0$ treating it as a system and treating the rest of the spins as a heat bath. 
For simplicity, we rewrite $H$ as $H=H_0+H_{\rm int}+H_B$, where 
\begin{equation}
H_0=\bd(t)\cdot \bsigma_0, \quad  H_{\rm int}=\sum_i J_{0i} \bsigma_0\cdot\delta \bsigma_i
\end{equation}
with $\delta \bsigma_i=\bsigma_i+\hat \bd_0(t)$, $\bd(t)=\hat\bd_0(t)(d-\sum_{i}J_{0i})$, and $H_B(t)$ denotes all the terms which are independent of $\bsigma_0$, i.e., describes the dynamics of the rest of the spins $\bsigma_i$, for $i>0$. 

The kinetic equation describing the dynamics of the total density matrix of the system in the interaction picture, $\tilde\rho$, reads~\cite{Breuer_Petruccione}
\begin{equation}
\partial_t \tilde\rho(t)=-\int_0^t dt' [\tilde H_{\rm int}(t),[\tilde H_{\rm int}(t'),\tilde\rho(t')]],
\label{eq:kineticRho}
\end{equation}
where $\tilde H_{\rm int}(t)=\hat U_B^{\dagger}(t)\hat U_0^{\dagger}(t)H_{\rm int}\hat U_0(t)\hat U_B(t)$ is the interaction picture representation of $H_{\rm int}$, and
\begin{equation}
    \hat U_0(t)=\cT e^{-i\int_0^t dt'H_0(t')} , ~ \hat U_B(t)=\cT e^{-i\int_0^t dt'H_B(t')}
\end{equation}
are the time evolution operators of the $i=0$ spin and the rest of the spins. We also define $\tilde\bsigma_i(t)=\hat U_B^{\dagger}(t)\hat U_0^{\dagger}(t)\bsigma_i\hat U_0(t)\hat U_B(t)$ to denote spins in the interaction picture.

Next, we focus on a single spin $\tilde\bsigma_0$, described by a reduced density matrix $\tilde\rho_0$, and assume the rest of the spins, described by the density matrix $\tilde \rho_B$, as its heat bath. This heat bath essentially describes vibrational modes of the cloud around an instantaneous ferromagentic state, in which all the spins are pointing along $-\bd_0(t)$. The correlation time $\tau_c$ of the vibrations decays with $J_{ij}$. Here, we consider the limit, $\omega \tau_c\ll1$, in which the vibrational modes  thermalize much faster than the change in $\bd_0(t)$.  Under these conditions,  the heat bath can be assumed in the instantaneous thermal equilibrium.

Furthermore, for large $N_{\rm c}$, we assume that $\tilde\rho_B$ is approximately unaffected by the interaction with $\tilde\bsigma_0$, therefore, we can write $\tilde\rho(t)=\tilde \rho_0(t)\otimes \tilde\rho_B(t)$.
Separating Eq.~\eqref{eq:kineticRho} to components dependent on $\tilde\bsigma_0$ and the heat bath, we arrive at
\begin{widetext}
\begin{equation}
\begin{split}
\partial_t \tilde\rho(t) = -\sum_{ij}J_{0i}J_{0j}\int_0^t dt'
&[\tilde\sigma^\alpha_0(t)\tilde\sigma^\beta_0(t')\tilde\rho_0(t')\delta\tilde\sigma^\alpha_i(t)\delta\tilde\sigma^\beta_j(t')\tilde\rho_B(t')
+\tilde\rho_0(t')\tilde\sigma^\beta_0(t')\tilde\sigma^\alpha_0(t)\tilde\rho_B(t')\delta\tilde\sigma^\beta_j(t')\delta\tilde\sigma^\alpha_i(t)-\\
-&\tilde\sigma^\beta_0(t')\tilde\rho_0(t')\tilde\sigma^\alpha_0(t)\delta\tilde\sigma^\beta_j(t')\tilde\rho_B(t')\delta\tilde\sigma^\alpha_i(t) -\tilde\sigma^\alpha_0(t)\tilde\rho_0(t')\tilde\sigma^\beta_0(t')\delta\tilde\sigma^\alpha_i(t)\tilde\rho_B(t')\delta\tilde\sigma^\beta_j(t')],
\end{split}
\end{equation}
where $\alpha, \beta=x,y,z$.
Next, we trace over the degrees of freedom of the heat bath, arriving at 
\begin{equation}
\begin{split}
\partial_t \tilde\rho_0(t)=-\sum_{ij}J_{0i}J_{0j}\int_0^t dt' 
[&(\tilde\sigma^\alpha_0(t)\tilde\sigma^\beta_0(t')\tilde\rho_0(t')-\tilde\sigma^\beta_0(t')\tilde\rho_0(t')\tilde\sigma^\alpha_0(t))\chi_{ij}^{>,\alpha\beta}(t,t')+\\
+&(\tilde\rho_0(t')\tilde\sigma^\beta_0(t')\tilde\sigma^\alpha_0(t)-\tilde\sigma^\alpha_0(t)\tilde\rho_0(t')\tilde\sigma^\beta_0(t'))
\chi_{ij}^{<,\alpha\beta}(t,t')],
\label{eq:Master2a}
\end{split}
\end{equation}
\end{widetext}
where 
\begin{align}
\chi^{>,\alpha\beta}_{ij}(t,t') &= {\rm Tr_B}\{ \delta\tilde\sigma^\alpha_i(t)\delta\tilde\sigma^\beta_j(t')\tilde\rho_B(t') \} ,\\ 
\chi^{<,\alpha\beta}_{ij}(t,t') &= {\rm Tr_B}\{ \delta\tilde\sigma^\beta_j(t')\delta\tilde\sigma^\alpha_i(t)\tilde\rho_B(t') \} ,
\end{align}
are components of the spin correlation function of the heat bath. 

To make progress, we use the Markov approximation setting $\tilde\rho_0(t')\approx \tilde\rho_0(t)$, which is valid in the regime $\lambda\tau_c\ll1$, where $\lambda$ is the characteristic rate of change of $\tilde\rho_0$, see below. Then we transform Eq.~\eqref{eq:Master2a} to the original basis by multiplying by $\hat U_0(t)$ from the left and $\hat U_0^\dagger(t)$ from the right. At this stage, we need to calculate $\hat U_0(t)\tilde\sigma_0^\alpha(t')\hat U^\dagger_0(t)$, which we do by expanding the evolution operator in the eigenbasis of $H_0(t)$, using $\hat U_0(t)=\sum_n e^{-i\varepsilon_n t} \ket{n(t)}\bra{n}$, where $\ket{n(t)}$ is the $n$-th eigenstate of $H_0(t)$ for $n=\pm$ at time $t$ and $\ket{n}$ is the same eigenstate at time $t=0$; $\varepsilon_n$ denotes the corresponding eigenenergy.
An explicit calculation yields
\begin{equation}
\hat U_0(t)\tilde\sigma_0^\alpha(t')\hat U^\dagger_0(t)=\sum_{mn} e^{-i(\varepsilon_m- \varepsilon_n)(t-t')}\sigma^\alpha_{0,mn}\ket{m}\bra{n},
\label{eq:Master2}
\end{equation}
where $\sigma^\alpha_{0,mn}=\langle{m}|\sigma_0^\alpha|{n}\rangle$, and  we used $\langle{m(t)}|{n(t')}\rangle\approx\delta_{mn}$ up to corrections of the order of $\sim \cO(\omega\tau_c)$.
After the transformation, Eq.~\eqref{eq:Master2a} reads 
\begin{equation}
\begin{split}
&\partial_t \rho_0(t)+i[H_0(t), \rho_0(t)]=\\
&-\lambda_{mn}^{>,\alpha\beta}(\sigma^\alpha_0\sigma^\beta_{0,mn}\rho_0(t)-\sigma^\beta_{0,mn}\rho_0(t)\sigma^\alpha_0) \\
&-\lambda_{mn}^{<,\alpha\beta}(\rho_0(t)\sigma^\beta_{0,mn}\sigma^\alpha_0-\sigma^\alpha_0\rho_0(t)\sigma^\beta_{0,mn})
,
\label{eq:Master3}
\end{split}
\end{equation}
where
\begin{equation}
\lambda_{mn}^{\gtrless,\alpha\beta}(t)=\sum_{ij}J_{0i}J_{0j} \int _0^{t} dt' e^{-i(\varepsilon_m-\varepsilon_n)(t-t')}\tilde\chi^{\gtrless,\alpha\beta}_{ij}(t,t').
\end{equation}

Finally, we transform Eq.~\eqref{eq:Master3} to the instantaneous basis of $H_0(t)$, using the transformation unitary $\hat U_I(t)$, such that
\begin{align}
\hat U_I^\dagger(t)\sigma^\alpha_i\hat U_I(t) &= \breve\sigma^\alpha_i(t) \\
\hat U_I^\dagger(t)\rho_i(t)\hat U_I(t) &= \breve\rho_i(t).
\end{align}
In the low-temperature and weak interaction limit, $\Delta\gg \sum_i |J_{ij}|$ and $\Delta\gg T$, $\langle \bsigma_i(t)\rangle\approx -\hat \bd(t)$ or alternatively $\langle \breve\bsigma_i(t)\rangle\approx -\hat \bz$, up to corrections $\sim \cO(J_{ij}/\Delta)$, $\sim \cO(\omega/\Delta)$, and $\sim \cO(T/\Delta)$, where $\Delta$ is the instantaneous gap of $H_0(t)$.
In this regime,
$\chi^{>,+-}_{ij}(t,t')$ 
is only non-zero for $\alpha=+$ and $\beta=-$, up to quadratic order corrections in small parameters. 
Similarly, $\chi^{<,-+}_{ij}(t,t')$ is also non-zero and the rest of the correlations can be neglected. 
We define 
\begin{equation}
    \chi^{\alpha\beta}_{ij}(\Delta t, \bar t)=\chi^{\alpha\beta}_{ij}(\bar t+\frac12\Delta t,\bar t-\frac12\Delta t),
\end{equation}
where $\Delta t=t-t'$ defines the fast scale proportional to $\tau_c$ and $\bar t=(t+t')/2$ defines the slow scale proportional to $\omega$. In the regime of local equilibrium, $\omega \tau_c\ll1$, the two scales separate, and we can further approximate $\bar t\approx t$.
Under the assumption of local equilibrium, the Fourier transformed correlation function, $\tilde\chi^{\alpha\beta}_{ij}(\Omega, t)=\int_{-\infty}^{\infty} d\Delta t e^{-i\Omega \Delta t}\chi^{\alpha\beta}_{ij}(\Delta t,t)$ is at thermal equilibrium at any $t$, satisfying  $\tilde\chi_{ij}^{>,-+}(\Omega, t)=\tilde\chi_{ji}^{<,+-}(-\Omega, t)$.

Finally, the r.h.s. of Eq.~\eqref{eq:Master3} can be written as
\begin{equation}
\begin{split}
&\lambda[2\breve\sigma^-_0\breve\rho_0(t)\breve\sigma^+_0-\breve\sigma^+_0\breve\sigma^-_0\breve\rho_0(t)-\breve\rho_0(t)\breve\sigma^+_0\breve\sigma^-_0],
\label{eq:Master4}
\end{split}
\end{equation}
where
$\lambda=\sum_{ij}J_{0i}J_{0j} \tilde\chi^{>,-+}_{ij}(\Delta,t)$, and we neglected the Lamb-shift term for simplicity. The expression in Eq.~\eqref{eq:Master4} vanishes for $\breve\rho_0=\breve\rho_{\rm eq}=\begin{pmatrix}
0 & 0 \\
0 & 1 
\end{pmatrix}$. Expanding $\breve\rho_0$ to the linear order in $\delta\breve\rho=\breve\rho_0-\breve\rho_{\rm eq}$, Eq.~\eqref{eq:Master4} transforms to 
\begin{equation}
    -\begin{pmatrix}
\frac{\delta\breve\rho_{11}}{\tau_1} & \frac{\delta\breve\rho_{12}}{\tau_2} \\
\frac{\delta\breve\rho_{21}}{\tau_2} & \frac{\delta\breve\rho_{22}}{\tau_1},
\end{pmatrix}
\end{equation}
where $\tau_1^{-1}=2\tau_2^{-1}=2\lambda$, and $\delta\breve\rho_{ij}$ are different components of $\delta\breve\rho$.


%